\documentclass[11pt]{amsart}
\numberwithin{equation}{section}
\numberwithin{figure}{section}
\numberwithin{table}{section}

\usepackage{macros}

\title{Mirror Symmetry and Level-rank Duality\\ for 3d $\mathcal{N} = 4$ Rank 0 SCFTs}
\author{Thomas Creutzig}
\address{Department of Mathematical and Statistical Sciences\\
	University of Alberta\\
	632 CAB, Edmonton, Alberta, Canada T6G 2G1}
\email[T.~Creutzig]{creutzig@ualberta.ca}

\author{Niklas Garner}
\address{Department of Physics\\
	University of Washington\\
	Seattle, WA 98195-1560, USA}
\email[N.~Garner]{nkgarner@uw.edu}

\author{Heeyeon Kim}
\address{Department of Physics\\ Korea Advanced Institute of Science and Technology\\ Daejeon 34141, Republic of Korea}
\email[H.~Kim]{heeyeon.kim@kaist.ac.kr}

\begin{document}

\begin{abstract}
	We introduce a family of 3d $\mathcal{N}=4$ superconformal field theories that have zero-dimensional Coulomb and Higgs branches and propose that the rational vertex operator algebras $W^{\textnormal{min}}_{k - \scriptstyle{\frac{1}{2}}}(\mathfrak{sp}_{2N})$ and $L_{k}(\mathfrak{osp}_{1|2N})$ model the modular tensor categories of line operators in their topological $A$ and $B$ twists, respectively. Our analysis indicates that the action of 3d mirror symmetry on this family of theories is related to a novel level-rank duality and leads to several conjectural $q$-series identities of independent interest.
\end{abstract}

\maketitle
\tableofcontents

\section{Introduction}

Three-dimensional (3d) topological quantum field theories (TQFTs) often admit chiral/holomorphic boundary conditions furnishing a vertex operator algebra (VOA) of boundary local operators, generalizing the connection between Chern-Simons gauge theories and chiral Wess-Zumino-Witten models \cite{Witten:1988hf, Elitzur:1989nr}.
The resulting interplay between 3d TQFTs and VOAs has been a rich source of insight into both subjects.
On one hand, the representation theory of a boundary VOA allows for an algebraic description of line operators in terms of modules for the boundary VOA, state spaces in terms of conformal blocks, and (sometimes) partition functions by way of the Reshetikhin-Turaev construction \cite{Reshetikhin:1991tc} or its non-semisimple generalization due to Costantino, Geer, and Patreau-Mirand \cite{Costantino:2012it}.
On the other, physical considerations coming from the bulk TQFT can lead to insights into the structure of a given boundary VOA as well as unexpected relations between seemingly disparate VOAs or different algebraic structures all together; examples of these insights include natural deformations coming from background fields \cite{Gaiotto:2016wcv, Feigin:2022nsb}, level-rank dualities between different VOAs \cite{Naculich:1990bu, Naculich:1990hg, Nakanishi:1990hj, MR3162483, Hsin:2016blu, Cordova:2018qvg, Creutzig:2019psu, Creutzig:2021ext}, and Kazhdan-Lusztig correspondences relating different descriptions of bulk line operators for the same TQFT \cite{Kazhdan1993, Feigin:2005xs, Feigin:2006iv, Creutzig:2023avg, Creutzig:2024abs}.

The 3d TQFTs of interest to this paper are realized as topological twists of QFTs with $\CN = 4$ supersymmetry.
Much like their 2d $\CN=(2,2)$ cousins \cite{Witten:1988xj}, 3d $\CN=4$ theories admit two topological twists: the $A$ twist, a dimensional reduction of the 4d Donaldson-Witten twist \cite{Witten:1988ze}, and the $B$ twist, first introduced by Blau and Thompson \cite{Blau:1996bx} and studied by Rozansky and Witten in the context of $\sigma$-models \cite{Rozansky:1996bq}.%
\footnote{The $B$ (resp. $A$) twist is sometimes called the $C$ (resp. $H$) twist, \emph{cf.} \cite{Gaiotto:2016wcv}, or the Rozansky-Witten (resp. twisted Rozansky-Witten) twist, \emph{cf.} \cite{Bullimore:2015lsa}.} %
The resulting TQFTs are cohomological in nature and, unlike most familiar 3d TQFTs, often admit non-trivial bulk local operators.
Physically, the algebras of these local operators in the $A$ and $B$ twists correspond to the algebras of holomorphic functions on the Coulomb and Higgs branches of vacua, respectively.
For the most commonly studied 3d $\CN=4$ theories, these branches of vacua are finite-dimensional varieties and, correspondingly, the algebras of local operators are infinite dimensional: the Higgs branch suffers from no quantum corrections \cite{Argyres:1996eh} and can be described as a complex symplectic or hyperk\"{a}hler reduction \cite{Hitchin:1986ea}, whereas the Coulomb branch suffers from both perturbative and non-perturbative corrections \cite{Seiberg:1996bs, Seiberg:1996nz} and lacked a precise definition until the work of Braverman, Finkelberg, and Nakajima \cite{Nakajima:2015txa, Braverman:2016wma}, see also \cite{Bullimore:2015lsa} for a related physical analysis.
Three-dimensional $\CN=4$ QFTs also enjoy a remarkable duality known as 3d mirror symmetry \cite{Intriligator:1996ex, deBoer:1996mp} that relates theories $\CT$ and $\CT^\vee$ in a way that exchanges their Coulomb and Higgs branches.
The equivalences of TQFTs
\be
    \CT^A \simeq (\CT^\vee)^B \hspace{2cm} \CT^B \simeq (\CT^\vee)^A
\ee
induced by 3d mirror symmetry make this class of QFTs ripe with mathematical insights; see \emph{e.g.} \cite{Webster2023} for a recent survey of the mathematics of 3d mirror symmetry, also known as symplectic duality \cite{Braden:2014iea, Bullimore:2016nji}.

Boundary VOAs for topologically twisted 3d $\CN=4$ theories first appeared in \cite{Gaiotto:2017euk, Costello:2018fnz} and have since seen a great deal of attention.
These boundary VOAs contain information about the geometry of Coulomb and Higgs branches \cite{Costello:2018swh, Beem:2023dub, Coman:2023xcq}, they bear witness to 3d mirror symmetry \cite{Ballin:2022rto, Ballin:2023rmt}, and they exhibit Kazhdan-Lusztig correspondences \cite{Creutzig:2024abs}; see \emph{e.g.} \cite{Creutzig:2021ext, Garner:2022rwe, Beem:2022vfz, Yoshida:2023wyt, Garner:2023pmt, Gang:2023rei, Gang:2023ggt, Ferrari:2023fez, Coman:2023xcq, Dedushenko:2023cvd, Baek:2024tuo, Gang:2024tlp} for an incomplete selection of other recent examples.
The VOAs describing topologically twisted 3d $\CN=4$ theories are generally logarithmic: they have modules that are reducible (the module has a non-trivial submodule) but indecomposable (the module cannot be written as a direct sum of two modules).
This is a necessity for these theories as bulk local operators in a cohomological TQFT are realized by self-extensions, \emph{i.e.} derived endomorphisms, of the vacuum module of a boundary VOA \cite{Costello:2018swh}.

More recently, there has been increased interest in 3d $\CN=4$ theories whose Coulomb and Higgs branches of vacua are both $0$-dimensional; these theories are said to be rank 0 \cite{Gang:2021hrd}.
The known examples of rank 0 theories come from many sources: they can come from dimensional reduction of 4d \cite{Dedushenko:2018bpp, Dedushenko:2023cvd} or 6d \cite{Choi:2022dju, Gang:2024tlp}, from coupling certain $\CN=4$ theories to Chern-Simons gauge fields \cite{Gang:2022kpe, Gang:2023ggt} via a mechanism discovered by Gaiotto and Witten \cite{Gaiotto:2008sd}, and there are many seemingly spurious examples, such as \cite{Gang:2018huc, Gang:2023rei, Baek:2024tuo}, coming from unexpected supersymmetry enhancement.
When the underlying 3d $\CN=4$ theory is rank $0$, the boundary VOAs describing its topological twists are highly constrained.
The vacuum module of such a boundary VOA must necessarily be both simple and have no non-trivial self-extensions in order to account for the absence of bulk local operators.
It is also expected that such a VOA must be $C_2$-cofinite (also called lisse) \cite{Ferrari:2023fez}.
In particular, based off analogy with the Higgs branch conjecture of Beem and Rastelli \cite{Beem:2017ooy} for VOAs coming from 4d $\CN=2$ superconformal field theories (SCFTs) as well as explicit examples \cite{Costello:2018swh, Coman:2023xcq} and free-field realizations \cite{Beem:2023dub}, the reduced part of the algebra of holomorphic functions on the associated variety of this boundary VOA is expected to realize the reduced part of the algebra of bulk local operators in the \emph{other} twist.
If the other twist also has no local operators, the 3d analog of the Higgs branch conjecture implies the associated variety must be a point and hence the boundary VOA must be $C_2$-cofinite.

In this paper we introduce a family of 3d $\CN=4$ rank 0 SCFTs $\CT_{N,k}$ for positive integers $N, k$ that extends the family of theories $\CT_k = \CT_{1,k}$ recently discovered by the third author \cite{Gang:2023rei}, see also \cite{Gang:2018huc} where the so-called minimal rank 0 SCFT $\CT_{\textnormal{min}} = \CT_{1,1}$ first appeared.
The SCFTs $\CT_{N,k}$ are realized in a similar fashion to \cite{Gang:2018huc, Gang:2023rei} by deforming a Lagrangian $\CN=2$ theory and flowing to the IR; see Section \ref{sec:TNk} for more details.
We also introduce a second family of 3d $\CN=4$ SCFTs $\ol{\CT}_{N,k}$ and propose a novel instance of 3d mirror symmetry for these rank 0 SCFTs.
\be
	\textnormal{3d mirror symmetry:} \qquad \CT_{N,k} \longleftrightarrow \ol{\CT}_{k,N} = (\CT_{N,k})^\vee
\ee
Boundary VOAs for topological twists of $\CT_k = \CT_{1,k}$ were recently proposed in \cite{Gang:2023rei} and \cite{Ferrari:2023fez}: the $A$ twist is modeled by the non-unitary minimal model $M(2,2k+3)$ and the $B$ twist is modeled by the simple affine VOA $L_k(\fosp_{1|2})$.
Due to the non-perturbative nature of the deformation used to realize $\CT_{N,k}$, at present it is not possible to directly verify these proposals and these boundary VOAs were deduced indirectly, \emph{e.g.} by computing (super)characters and modular data by way of supersymmetric localization.
In Sections \ref{sec:osp} and \ref{sec:Wmin} we extend this analysis to $\CT_{N,k}$, leading us to the following proposal: line operators are modeled by modules for
\be
	\CV^A_{N,k} = W^{\textnormal{min}}_{k-\scriptstyle{\frac{1}{2}}}(\fsp_{2N}) 
\ee
in the $A$ twist and
\be	
	\CV^B_{N,k} = L_k(\fosp_{1|2N})
\ee
in the $B$ twist.
Finally, in Section \ref{sec:mirrorBC} we consider the mirror boundary conditions for $\ol{\CT}_{k,N}$ and describe consequences thereof.

In the remainder of this introductory section we illustrate two interesting mathematical results indicated by our analysis.
The first is a collection of conjectural refined $q$-series identities coming from the various physical realizations of these VOAs and their modules.
The second consequence we describe is a collection of level-rank dualities that come from our mirror symmetry analysis.
We end with some further questions raised by our analysis.

\subsection{Character formulas}
An immediate consequence of our proposals is a collection of conjectural refined $q$-series identities coming from the mirror realizations of the VOAs $W^{\textnormal{min}}_{k-\scriptstyle{\frac{1}{2}}}(\fsp_{2N})$ and $L_k(\fosp_{1|2N})$.

Starting with the affine VOA $L_k(\fosp_{1|2N})$, we find two conjectural realizations for its normalized vacuum character.
First, there is its realization as a Dirichlet half-index of $\CT_{N,k}$:
\be
\label{conj:ospDir}
\begin{aligned}
	\ol{\chi}[L_k(\fosp_{1|2N})](x_\alpha;q) & = \Tr_{L_k(\fosp_{1|2N})} (-1)^F q^{L_0} \prod_{\alpha = 1}^{N} x_\alpha^{h_{\alpha,0}}\\
	& \overset{?}{=} \sum\limits_{m_{ia}}{}' q^{\scriptstyle{\frac{1}{2}} \sum\limits_{i,j =1}^{k}\sum\limits_{a,b=1}^{2N-1} T_{ij} C_{ab}m_{ia}m_{jb}}\frac{\prod\limits_{\alpha=1}^N (q^{1-m_{k(2\alpha-1)}} x_{\alpha}^{-1})_\infty x_{\alpha}^{\sum\limits_{j=1}^{k} \sum\limits_{b=1}^{2N-1} jC_{(2\alpha-1) b} m_{jb}}
	}{(q)_\infty^N \left(\prod\limits_{i=1}^{k-1} \prod\limits_{a=1}^{2N-1}(q)_{-m_{ia}}\right) \left(\prod\limits_{\beta =1}^{N-1}(q)_{-m_{k (2\beta)}}\right)}
\end{aligned}
\ee
where the $h_{\alpha,0}$ are generators of a Cartan subalgebra of $\fsp_{2N}$ and $\sum'$ refers to a summation with $m_{ia}$ a non-positive integer if $i < k$ or if $a$ is even, and any integer otherwise.
In this expression, $C_{ab}$, $a,b = 1, \dots, 2N-1$, denotes the Cartan matrix of $A_{2N-1}$ and we use the notation $(x)_\infty = \prod_{n \geq 0} (1-x q^n)$; we also denote $(x)_m = (x)_\infty /(x q^m)_\infty$.
It was shown in \cite{Ferrari:2023fez} that this equality is true to all orders in $q$ for $k = N = 1$ and verified to order $q^{10}$ for $N = 1$ and $k \leq 3$; we have verified that this generalized form continues to hold for all $N$, $k$ with $N + k \leq 4$ up to at least order $q^5$.
An expression for the vacuum character of this form is particularly noteworthy because when we specialize the fugacities/Jacobi variables $x_\alpha$ to $1$ it naturally leads to a Nahm (or fermionic) sum:
\be
\label{conj:ospNahm}
	\ol{\chi}[L_k(\fosp_{1|2N})](1;q) \overset{?}{=} \sum_{n_{ia} \geq 0} \frac{q^{\scriptstyle{\frac{1}{2}} \sum\limits_{i,j =1}^{k}\sum\limits_{a,b=1}^{2N-1} T_{ij} C_{ab}n_{ia}n_{jb}}}{(q)_{n_{ia}}}
\ee
where $T_{ij} = \min(i,j)$, $i,j = 1, \dots, k$, is the inverse of the Cartan matrix of the rank $k$ tadpole diagram; we also set $n_{ia} = -m_{ia}$.
Expressions for VOA characters as Nahm sums have a long history that stems from the existence of massive integrable deformations of 2d CFTs and their relations to the Bethe ansatz \cite{Zamolodchikov:1978xm, Zamolodchikov:1989hfa, Zamolodchikov:1989cf, Reshetikhin:1989qg}.
These expressions are also naturally tied to dilogarithm identities \cite{Nahm:1992sx, Zagier:2007knq} and arise in the related context of the 3d-3d correspondence \cite{Cheng:2022rqr}.

We also note that Nahm sums play a fundamental role in Rogers-Ramanujan-like identities \cite{Kedem:1993wt, Berkovich:1997ht} and Andrews-Gordon-like identities \cite{Warnaar2012, Bringmann2014}. In fact, Eq. \eqref{conj:ospNahm} is precisely the identity in Conjecture 1.1 of \cite{Warnaar2012} (with $p = k$), hence Eq. \eqref{conj:ospDir} can be viewed as a refinement of this conjecture.
The first two authors prove that this Andrews-Gordon-like identity holds for all $N$, $k$ in \cite{Creutzig:2025tgh} using a variant of a construction due to Feigin and Stoyanovsky \cite{Feigin:1993qr, Stoyanovsky1998}, where an analogous Nahm sum with $(C_{ab})$ replaced by the Cartan matrix of $A_{2N}$ was equated to the (normalized) vacuum character of $L_{k}(\fsp_{2N})$.

A second expression for the normalized vacuum character of $L_k(\fosp_{1|2N})$ comes from its realization as a Neumann half-index of $\ol{\CT}_{k,N}$:
\be
\label{conj:ospNeu}
\begin{aligned}
	\ol{\chi}[L_k(\fosp_{1|2N})](x_\alpha;q) & \overset{?}{=} \oint \prod\limits_{\alpha=1}^N \prod\limits_{m=1}^{2k-1}\bigg(\frac{\textnormal{d} z_{\alpha m}}{2\pi i \, z_{\alpha m}} \frac{(q)_\infty}{(z_{\alpha m})_\infty} \bigg)\\
	& \times \prod\limits_{\alpha=1}^{N}\prod\limits_{\mu=1}^{k}\left(FF\bigg(x_\alpha \prod\limits_{\beta=\alpha}^N \frac{z_{\beta (2\mu-1)}}{z_{\beta(2\mu-2)}}\bigg) FF\bigg(x_\alpha^{-1} \prod\limits_{\beta=\alpha}^N \frac{z_{\beta (2\mu-1)}}{z_{\beta(2\mu)}}\bigg)\right)\\
\end{aligned}
\ee
where we set $z_{\alpha0} = z_{\alpha(2k)} = 1$ for all $\alpha$, $FF(x) = (x)_\infty (q x^{-1})_\infty$ is the vacuum character of two copies of the fermion VOA $F$ (rather, of a $bc$ ghost system), and the contour integral is taken over the product of unit circles $|z_{\alpha m}| = 1$.
We can explicitly show this equality is true to all orders in $q$ for $N = k = 1$, and have verified it to at least in $q^4$ for all $N$,$k$ with $N+k \leq 4$.

The normalized vacuum character of the $W$-algebra $W^{\textnormal{min}}_{k - \scriptstyle{\frac{1}{2}}}(\fsp_{2N})$ also admits expressions of the above form.
Its realization as a Dirichlet half-index leads to the following expression:
\be
\label{conj:WminDir}
\begin{aligned}
	\ol{\chi}[W^{\textnormal{min}}_{k - \scriptstyle{\frac{1}{2}}}(\fsp_{2N})](w_\nu;q) & = \Tr_{W^{\textnormal{min}}_{k - \scriptscriptstyle{\frac{1}{2}}}(\fsp_{2N})} q^{L_0} \prod_{\nu=1}^{N-1} w_\nu^{h_{\nu+1, 0}}\\
	&  \overset{?}{=} \sum\limits_{m_{ia}}{}'' q^{\scriptstyle{\frac{1}{2}} \sum\limits_{i,j =1}^{k}\sum\limits_{a,b=1}^{2N-1} T_{ij} C_{ab}m_{ia}m_{jb} + \sum\limits_{i=1}^k \sum\limits_{a=1}^{2N-1}(-1)^a i m_{ia} + \scriptstyle{\frac{1}{2}}\sum\limits_{j=1}^k\sum\limits_{\nu=1}^{N-1}j C_{(2\nu + 1)b}m_{jb}}\\
	& \qquad \times \frac{\prod\limits_{\nu=1}^{N-1} (q^{\scriptstyle{\frac{1}{2}}-m_{k(2\nu+1)}} w_{\nu}^{-1})_\infty w_\nu^{\sum\limits_{j=1}^k\sum\limits_{\nu=1}^{N-1}j C_{(2\nu + 1)b}m_{jb}}}{(q)_\infty^{N-1} \left(\prod\limits_{i=1}^{k-1} \prod\limits_{a=1}^{2N-1}(q)_{-m_{ia}}\right) \left((q)_{-m_{k 1}}\prod\limits_{\beta=1}^{N-1}(q)_{-m_{k (2\beta)}}\right)}
\end{aligned}
\ee
where $\sum''$ refers to a summation where $m_{ia}$ is a non-positive integer if $i < k$, if $a$ is even, or if $a = 1$, and any integer otherwise.
That this equality holds for $N = 1$, where $W^{\textnormal{min}}_{k-\scriptstyle{\frac{1}{2}}}(\fsp_{2}) \simeq M(2,2k+3)$, follows from the fermionic sum representations of $M(2, 2k+3)$ characters \cite{Feigin:1991wv, Nahm:1992sx, Kedem:1993ze, Berkovich:1994es, Nahm:1994vas}.
Note that specializing the fugacities/Jacobi variables $w_\nu \to q^{-\scriptstyle{\frac{1}{2}}}$ leads to the other Nahm sum appearing in Conjecture 1.1 and Theorem 1.2 of \cite{Warnaar2012}.
The realization of this vacuum character as a Neumann half-index leads to the following expression:
\be
\label{conj:WminNeu}
\begin{aligned}
	\ol{\chi}[W^{\textnormal{min}}_{k-\scriptstyle{\frac{1}{2}}}(\fsp_{2N})](w_\nu;q) & \overset{?}{=} \oint \prod\limits_{\alpha=1}^N \prod\limits_{m=1}^{2k-1}\bigg(\frac{\textnormal{d} z_{\alpha m}}{2\pi i \, z_{\alpha m}} \frac{(q)_\infty}{(z_{\alpha m})_\infty} \bigg)\\
	& \times \prod\limits_{\mu=1}^{k}\Bigg(FF\bigg(\prod\limits_{\beta=1}^N \frac{z_{\beta (2\mu-1)}}{z_{\beta(2\mu-2)}}\bigg) FF\bigg(q\prod\limits_{\beta=1}^N \frac{z_{\beta (2\mu-1)}}{z_{\beta(2\mu)}}\bigg)\\
	& \quad \times \prod\limits_{\nu=1}^{N-1}FF\bigg(q^{\scriptstyle{\frac{1}{2}}}w_\nu \prod\limits_{\beta=\nu+1}^N \frac{z_{\beta (2\mu-1)}}{z_{\beta(2\mu-2)}}\bigg) FF\bigg(q^{\scriptstyle{\frac{1}{2}}} w_\nu^{-1} \prod\limits_{\beta=\nu+1}^N \frac{z_{\beta (2\mu-1)}}{z_{\beta(2\mu)}}\bigg)\Bigg)
\end{aligned}
\ee
That this reproduces the vacuum character of $M(2,5)$ for $N = k = 1$ was established to all orders in $q$ in \cite{Ferrari:2023fez}.
We have checked both of these expressions agree with the normalized vacuum character $\ol{\chi}[W^{\textnormal{min}}_{k-\scriptstyle{\frac{1}{2}}}(\fsp_{2N})](w_\nu;q)$ to at least $q^4$ for all $N$, $k$ with $N + k \leq 4$.

There are analogous Neumann and Dirichlet expressions for characters of all modules of these VOAs.
When $N = 1$, the Dirichlet form of these characters are given in \cite{Gang:2023rei} and \cite{Ferrari:2023fez}, the former being related to the above fermionic sum representations of $M(2,2k+3)$ characters.
We extend these proposals to closed-form expressions for all modules when $k = 1$ for any $N$ and when $N = k = 2$ below, but we are unable to extend these examples to a proposal for general $N$, $k$.
One observation we make is that the form of these half-indices suggests a relation between specializations of these characters:
\be
	\ol{\chi}[L_k(\mu)](1;q) \overset{?}{=} \ol{\chi}[W^{\textnormal{min}}_{k-\scriptstyle{\frac{1}{2}}}(\mu^\sharp)](q^{-\scriptstyle{\frac{1}{2}}};q)
\ee
where $\mu^\sharp$ is the complementary partition to $\mu$, viewing the latter as a partition of $|\mu|$ that fits inside an $N \times k$ rectangle.
This proposal extends that of \cite{Ferrari:2023fez}, whereby $M(2,2k+3)$ characters are realized by specializing characters of $L_k(\fosp_{1|2})$.

\subsection{Level-rank duality and mirror symmetry}
\begin{figure}[h]
	\centering
	\begin{tikzpicture}
		\draw (1.75,1.25) -- (-2.25,1.25);
		\draw (0,1.5) node {$\CW_1$};
		
		\draw (2.25,0.5) -- (-2,0.5);
		\draw (0,0.75) node {$\CW_2$};
		
		\draw (2,-0.25) -- (-1.75,-0.25);
		\draw (0,0) node {$\CW_3$};
		
		\draw (2.5,1) -- (2.5,-1) -- (1.5,0) -- (1.5,2) -- cycle;
		\draw (1.75,-0.75) node {$\CB_R$};
		\draw[fill=white] (-1.5,1) -- (-1.5,-1) -- (-2.5,0) -- (-2.5,2) -- cycle;
		\draw (-2.25,-0.75) node {$\CB_L$};
		
		\draw[dashed] (1.75,1.25) -- (-2.25,1.25);
		\draw (1.75,1.25) node {$\bullet$};
		\draw (-2.25,1.25) node {$\bullet$};
		
		\draw[dashed] (2.25,0.5) -- (-1.75,0.5);
		\draw (2.25,0.5) node {$\bullet$};
		\draw (-1.75,0.5) node {$\bullet$};
		\draw[dashed] (2,-0.25) -- (-2,-0.25);
		\draw (2,-0.25) node {$\bullet$};
		\draw (-2,-0.25) node {$\bullet$};
	\end{tikzpicture}
	\caption{An illustration of the sandwich formed by boundary conditions $\CB_{L,R}$ with line operators $\CW_i$ stretched between them. The junction of the line operator $\CW_i$ with the boundary condition $\CB_{L,R}$ is identified with the module $\mathbf{M}_{L,R}[\CW_i]$ for the algebra $\CV_{L,R}$ of local operators on $\CB_{L,R}$.}
	\label{fig:sandwich}
\end{figure}
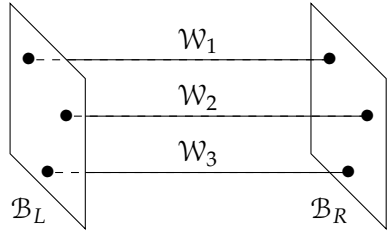

The second consequence we describe is a novel level-rank duality we deduce from our analysis and then prove using the branching rules found in \cite{Creutzig:2024uid}.
Chern-Simons gauge theories exhibit level-rank dualities that relate theories with different gauge groups and levels to one another \cite{Naculich:1990hg, Naculich:1990bu, Nakanishi:1990hj}, see also \cite{MR3162483, Hsin:2016blu}.
One interpretation of these dualities comes from the observation that these Chern-Simons gauge theories admit two different boundary VOAs $\CV_L$, $\CV_R$, one on a \emph{left} boundary and the other on a \emph{right} boundary, \emph{cf.} \cite{Creutzig:2021ext}.
For example, $SU(N)_k$ Chern-Simons theory has a right Dirichlet boundary condition furnishing the affine VOA $\CV_R = L_k(\fsl_N)$ as well as a left Neumann boundary condition furnishing the coset/commutant $\CV_L = \textnormal{Com}(L_k(\fsl_N), F^{2Nk}) \simeq L_{N}(\fgl_k)$, where $F^{2Nk}$ denotes the VOA of $2Nk$ free real fermions.
As the latter VOA also arises on a \emph{right} boundary of $U(k)_{N,N}$ Chern-Simons theory, it is possible to conclude the physical duality $SU(N)_k \simeq U(k)_{-N,-N}$, where the negation of the levels on the latter accounts for the requisite parity transformation.
That both of these VOAs model line operators of the same TQFT implies that their categories of modules must be equivalent, up to a reversal of the braiding due to the different orientations of the boundaries.
The equivalence is realized by identifying modules living at opposite ends of the same line operator; see Fig. \ref{fig:sandwich} for an illustration.

The existence of this braid-reversed equivalence of module categories for $\CV_L$ and $\CV_R$ implies it is possible to form a third VOA $\CV$ by gluing them along this equivalence \cite{Creutzig:2017anl, Creutzig:2019psu}.
The resulting VOA necessarily has a trivial category of modules and contains $\CV_L \otimes \CV_R$ as a subalgebra. Physically, the VOA $\CV$ obtained by gluing $\CV_L$, $\CV_R$ realizes the algebra of local operators in an effective 2d holomorphic QFT obtained by shrinking the size of the transverse direction.
Local operators in the effective 2d theory can be realized as local operators on either of the boundaries, giving rise to the product $\CV_L \otimes \CV_R$, together with all possible configurations of line operators stretched between the two boundaries, leading to the extension.
In the above Chern-Simons example, $L_k(\fsl_N)$ and $L_N(\fgl_k)$ are glued together to form the free-fermion VOA $F^{2Nk}$. For other examples of this interval reduction see \emph{e.g.} \cite{Sugiyama:2020uqh, Bullimore:2021rnr, Dedushenko:2021mds, Alekseev:2022gnr, Dedushenko:2022uay, Dedushenko:2023cvd}; see also \cite{Creutzig:2021ext} for examples of logarithmic level-rank dualities arising from twisted $\CN=4$ theories.
The converse is true too (modulo some technical assumptions \cite{Creutzig:2019psu}, see also \cite{mcrae2023general}): if VOAs $\CV_1$, $\CV_2$ are mutual cosets/commutants of one another inside a third VOA $\CV$ and if $\CV$ has a trivial category of modules, then the categories of modules for $\CV_1$ and $\CV_2$ realize equivalent braided-tensor categories up to a reversal of the braiding.

In Section \ref{sec:sandwich} we argue that the topologically twisted theories $\CT^A_{N,k}$ and $\CT^B_{N,k}$ each realize level-rank dualities of this form.
The key observation (see Section \ref{sec:3dmirror}) is that the mirror theory $\ol{\CT}_{k,N} = (\CT_{N,k})^\vee$ differs from $\CT_{k,N}$ by a parity transformation.
In particular, the \emph{right} boundary condition of $\ol{\CT}_{N,k}$ furnishing $L_N(\fosp_{1|2k})$ in the $A$ twist can equivalently be viewed as a \emph{left} boundary condition of $\CT_{N,k}$ furnishing $L_N(\fosp_{1|2k})$ in the $A$ twist.
There is an analogous statement for the $B$ twist.
In both cases, we are in the setup described above: a TQFT (say, $\CT^A_{N,k}$) with left and right boundary conditions furnishing VOAs.
We are thus lead to the following level-rank duality of VOAs.
\be
	\textnormal{level-rank duality:} \qquad W^{\textnormal{min}}_{k-\scriptstyle{\frac{1}{2}}}(\fsp_{2N}) \longleftrightarrow L_N(\fosp_{1|2k})
\ee
We prove this level-rank duality in Section \ref{sec:level-rank} by showing these VOAs are mutual commutatnts in the free-fermion VOA $F^{4Nk}$.

\subsection{Open questions and future directions}
In addition to the above conjectural character identities, our work leaves open several questions and directions for future study.

\subsubsection{Classical freeness}

VOAs arise in many corners of higher-dimensional SCFT, but it is still unclear what structural properties to expect these VOAs to satisfy.
One common feature is that of classical freeness, in the sense of van Ekeren and Heluani \cite{EH21}, but a good physical explanation for this is still missing.

Classical freeness means the following.
Let $\CV$ be a vertex algebra and let $R_\CV$ be its commutative Zhu algebra.
The associated scheme of $\CV$ is $X_\CV =$ Spec $R_\CV$ \cite{Ara15} and its reduced scheme is called the associated variety of $\CV$.
When $\CV$ arises from a 4d $\CN=2$ SCFT via the mechanism of \cite{Beem:2013sza}, the associated variety is conjectured \cite{Beem:2017ooy} to reproduce the Higgs branch of the 4d SCFT; when $\CV$ arises at the boundary of the $A$ (resp. $B$) twist of a 3d $\CN=4$ theory, then the associated variety conjecturally reproduces the Higgs (resp. Coulomb) branch \cite{Beem:2023dub, Ferrari:2023fez, Coman:2023xcq}.
The commutative Zhu algebra $R_\CV$ is the grade zero component of a commutative Poisson vertex algebra, realized as the associated graded $\textnormal{gr}(\CV)$ of $\CV$ (with respect to the Li filtration) \cite{Li2005}.
One should think of $\CV$ as a quantization of this commutative Poisson vertex algebra.
Let $(R_\CV)_\infty$ be the affine coordinate ring of the arc space $J_\infty(X_\CV)$, then there is always a surjection from $(R_\CV)_\infty$ onto $\textnormal{gr}(\CV)$ and the vertex algebra $\CV$ is called classically free if this surjection is in fact an isomorphism.
In summary, a vertex algebra is classically free if it is a quantization of the arc space of its associated scheme.

The simple affine VOAs $L_k(\fosp_{1|2N})$ at positive integer level $k$ are classically free by the results of \cite{Creutzig:2022sfx}. 
We also believe that $W^{\textnormal{min}}_{k-\scriptstyle{\frac{1}{2}}}(\fsp_{2N})$ are precisely the minimal $W$-algebras of type $\fsp_{2N}$ at admissible level that are classically free; this is indeed true for $N=1$ \cite{EH21}.
It is worth noting that the recent papers \cite{Baek:2024tuo, Gang:2024tlp} propose 3d $\CN = 4$ theories that admit boundary VOAs that are not classically free: Theorem 15.13 of \cite{EH21} shows that the only classically free minimal models are the $M(2,2k+3)$ and Theorem 6.1 of \cite{Li2021} shows that the only classically free $N=1$ minimal models are the $SM(2,4r)$.

\subsubsection{Other Dynkin types}

One of the central pieces of data that goes into the definition of our 3d $\CN=4$ theories is a pair of Dynkin diagrams: the Chern-Simons levels $\kappa$ used to define the theory (that deforms to) $\CT_{N,k}$ takes the form
\be
	\kappa = C(T_k)^{-1} \otimes C(A_{2N-1})\,,
\ee
where $C(X)$ denotes the Cartan matrix of a diagram $X$, \emph{cf.} Eq. \eqref{conj:ospNahm}.
More generally, Nahm sums of the form in Eq. \eqref{conj:ospNahm} are expected, \emph{cf.} \cite{Gliozzi:1994cs, Nahm:2004ch}, to have nice modular properties when we replace $T_k$ and $A_{2N-1}$ by any other ADET Dynkin diagrams, but it is presently unclear what boundary VOAs arise in these more general cases and which cases exhibit supersymmetry enhancement upon deformation.

It is worth noting that there are VOAs whose vacuum characters take the above form with $A_{2N-1}$ replaced by any simply-laced Dynkin diagram: the principal subspaces of Feigin and Stoyanovsky \cite{Feigin:1993qr}.
Based on the analyses of this paper and the work \cite{Creutzig:2025tgh}, we view this as merely an incidence of vacuum characters of two different VOAs, \emph{cf.} related work of Stoyanovsky \cite{Stoyanovsky1998} for type $A_{2N}$.
Indeed, as mentioned above, the examples with $A_{2N-1}$ replaced by $A_{2N}$ do not lead to $\CN=4$ theories but nonetheless can be deformed to gapped theories that flow to $USp(2N)$ Chern-Simons theories in the infrared.
We also note that one of the families of examples in \cite{Baek:2024tuo} giving rise to the $N=1$ minimal models $SM(2,4r+4)$ are realized from taking $\kappa = C(T_{2r})^{-1} \otimes C(T_1)$.

\subsubsection{First-principles derivation of boundary VOAs}

As mentioned above, we do not arrive at our proposed boundary VOAs directly, instead relying on indirect evidence to check the consistency of our proposal.
The same was true for the $N=1$ cases studied in \cite{Gang:2023rei, Ferrari:2023fez}, except for the $B$ twist of $\CT_{\textnormal{min}} \simeq \CT_{1} \simeq \CT_{1,1}$ where \cite{Ferrari:2023fez} presented direct evidence for the appearance of $L_1(\fosp_{1|2})$.
The approach advocated for in \cite{Ferrari:2023fez} was to realize these topological twists as deformations of an intermediate holomorphic-topological ($HT$) twist, which is available to any theory with $\CN\geq2$ supersymmetry and, in particular, the UV abelian gauge theory that is deformed to $\CT_{N,k}$.
This has proven to be a useful organizational principle for twisting highly supersymmetric theories and has applicability well beyond 3d $\CN=4$ theories, notably including the Langlands twists of 4d $\CN=4$ super Yang-Mills theories \cite{Elliott:2015rja} where this perspective was originally described.

In \cite{Ferrari:2023fez}, the affine VOA $L_1(\fosp_{1|2})$ was realized by adapting the techniques introduced in \cite{Costello:2020ndc} for studying the algebras of local operators on boundary conditions in $HT$-twisted gauge theories.
Tools for studying more general theories such as $\CT_{N,k}$, namely theories with superpotentials involving general monopole operators, are presently unavailable.
Once these tools have been developed, we believe that the examples studied in this paper should be relatively tractable.

Concretely, we expect that both of the boundary VOAs 
\be
	\CV^A_{N,k} = W^{\textnormal{min}}_{k-\scriptstyle{\frac{1}{2}}}(\fsp_{2N}) \qquad \textnormal{and} \qquad \CV^B_{N,k} = L_k(\fosp_{1|2N})
\ee
can be realized by deforming a single vertex algebra $\CV_{N,k}$ equipped with a 1-parameter family of quasi-conformal structures, with the quasi-conformal structure on $\CV^A_{N,k}$ and $\CV^B_{N,k}$ induced by their conformal vectors being compatible with those of $\CV_{N,k}$ at parameters (say) $\nu = -1$ and $\nu = 1$, respectively, \emph{cf.} \cite{Gang:2023rei, Ferrari:2023fez}.
Ignoring the quasi-conformal structure, one reasonable possibility for the vertex algebra $\CV_{N,k}$ is as a large-level limit of a deformable family \cite{Creutzig:2012sf, Creutzig:2014lsa} of vertex algebras containing $L_k(\fosp_{1|2N})$ in which the fermionic generators have regular OPEs with one another.%
\footnote{The large-level limit we have in mind corresponds to rescaling the fermionic generators by a factor of $\lambda \in \C^\times$ and then taking the $\lambda \to \infty$ limit.} %
Deforming the vertex algebra $\CV_{N,k}$ to the $B$ twist should be realized as in \cite{Ferrari:2023fez} by reintroducing the OPEs of these generators, whereas deforming to the $A$ twist should involve introducing a suitable differential.

The family of quasi-conformal structures on $\CV_{N,k}$ comes from spectral flow with respect to an outer $\C^\times[\![z]\!]$ symmetry of this vertex algebra.
This $\C^\times[\![z]\!]$ action extends an action of $\C^\times$ on the Lie algebra of zero modes: on the bosonic subalgebra the $\C^\times$ weights are given by the weights of $\frac{1}{2} \sum h_\alpha$ whereas on the fermionic subspace it gives $\theta_a$ weight $1$ if $a$ is odd and weight $0$ if $a$ is even.
For example, when $N = 1$ this gives $h$ and $\theta_-$ weight $0$, $e$ and $\theta_+$ weight $1$, and $f$ weight $-1$.
Physically, this $\C^\times$ action is induced by the action of a residual $R$-symmetry in the $HT$ twist (called $U(1)_S$ below) and the different quasi-conformal structures correspond to different choices of twisting homomorphism.
That this $\C^\times$ action extends to an action of $\C^\times[\![z]\!]$ is an expected feature of symmetries in $HT$-twisted theories \cite{Garner:2023wrc}.

\subsection*{Acknowledgments}
We would like to thank Dongmin Gang and Seungjoo Baek for useful discussions and collaborations on related work. 
The work of HK is supported by the Ministry of Education of the Republic of Korea and the National Research Foundation of Korea grant NRF-2023R1A2C1004965. 
NG is supported by funds from the Department of Physics and the College of Arts \& Sciences at the University of Washington.

\section{The theories \texorpdfstring{$\CT_{N,k}$}{TNk}}
\label{sec:TNk}

In this section we introduce a family of 3d $\CN=2$ Chern-Simons-matter theories and argue that they flow  to rank 0 $\CN=4$ SCFTs $\CT_{N,k}$ in the infrared after introducing a suitable superpotential, extending the family of theories $\CT_k = \CT_{1,k}$ studied in \cite{Ferrari:2023fez, Gang:2023rei}.
We then describe various aspects of the topologically twisted theories $\CT^A_{N,k}$ and $\CT^B_{N,k}$ coming from purely bulk considerations and propose a 3d mirror of $\CT_{N,k}$.

\subsection{$\CN=2$ Lagrangian description}
We consider $k(2N-1)$ abelian vector multiplets, each coupled to a single charge 1 chiral multiplet, coupled to one another through effective Chern-Simons levels of the form
\be
\label{eq:CSlevel}
	\kappa = C(T_k)^{-1} \otimes C(A_{2N-1})\,,
\ee
where $C(T_k)$, $C(A_{2N-1})$ are the Cartan matrices of the tadpole diagram $T_k$, obtained by folding the $A_{2k}$ diagram, and the $A_{2N-1}$ Dynkin diagram, respectively.
Explicitly, the matrix elements of $C(T_k)^{-1} = (T_{ij})$ and $C(A_{2N-1}) = (C_{ab})$ are given by
\be
	T_{ij} = \min(i,j) \qquad \textnormal{and} \qquad  C_{ab} = 2 \delta_{ab} - \delta_{a (b+1)} - \delta_{a (b-1)}\,.
\ee

This theory does not have the $\CN=4$ supersymmetry necessary for topological twisting, but we expect that it will flow to one in the IR after we deform by a suitable superpotential $W_{N,k}$.
For $N = 1$, this superpotential is given by a sum of gauge-invariant $\frac{1}{2}$-BPS bare monopole operators of the form \cite{Gang:2023rei}
\be
	W_{1,k} = \sum_{i=1}^{k-1} V_{\mathfrak{m}_i} \qquad (\mathfrak{m}_{i})_j = 2 \delta_{ij} - \delta_{(i-1)j} - \delta_{(i+1)j}\,.
\ee
There are no gauge-invariant \emph{bare} monopole operators when $N > 1$, but we find a similarly simple collection of $\frac{1}{2}$-BPS dressed monopole operators when $k = 1$:
\be
	W_{N, 1} = \sum_{a=1}^{2N-2} \phi_{a-1} \phi_{a+2} V_{\mathfrak{m}_a} \qquad (\mathfrak{m}_a)_b = \delta_{ab} + \delta_{(a+1)b}
\ee
where we set $\phi_{0} = \phi_{2N} = 1$.
More generally, we find $k(2N-1)-1$ monopole operators coming in two types that we can use to deform the above UV theory.
The first $(k-1)(2N-1)$ monopole operators generalize those appearing in $W_{1,k}$ and take the form
\be
	\phi_{i(a-1)} \phi_{i(a+1)} V_{\mathfrak{m}_{ia}} \qquad (\mathfrak{m}_{ia})_{jb} = (2\delta_{ij}-\delta_{(i-1) j} - \delta_{(i+1)j})\delta_{ab}
\ee
for $i = 1, \dots, k-1$ and $a = 1, \dots, 2N-1$, where $\phi_{i 0} = \phi_{i (2N)} = 1$.
The remaining $2N-2$ monopole operators generalize those appearing in $W_{N,1}$ and take the form
\be
	\phi_{k(a-1)}\phi_{k(a+2)} V_{\mathfrak{m}_a} \qquad (\mathfrak{m}_a)_{jb} = (\delta_{jk}-\delta_{j(k-1)})(\delta_{ab} + \delta_{(a+1)b})
\ee
for $a = 1, \dots, 2N-2$.
The full superpotential is then a sum of these $k(2N-1)-1$ monopole operators:
\be
	W_{N,k} = \sum_{i=1}^{k-1} \sum_{a=1}^{2N-1} \phi_{i(a-1)} \phi_{i(a+1)} V_{\mathfrak{m}_{ia}} + \sum_{a=1}^{2N-2} \phi_{k(a-1)}\phi_{k(a+2)} V_{\mathfrak{m}_a}\,.
\ee
Deforming by this monopole superpotential, the original global symmetry $U(1)^{(2N-1)k}\times U(1)_R$ breaks to $U(1)_S\times U(1)_R$, where $U(1)_S$ is generated by
\be
	S = \sum\limits_{a=1}^{2N-1}\sum\limits_{i=1}^{k} (-1)^{a+1} i M_{ia}\,.
\ee
Here $M_{ia}$ is the generator of the topological symmetry for the $U(1)_{ia}$ factor of the gauge group.

\subsection{Evidence for 3d $\CN=4$ supersymmetry}
As argued in \cite{Gang:2018huc} for the theory $\CT_{1,1}$ and \cite{Gang:2023rei} for the family $\CT_{1,k}$, we expect the theories $\CT_{N,k}$ described above exhibit enhancement to $\CN=4$ supersymmetry in the infrared.
Our main piece of evidence for this enhancement comes from the superconformal index, which is simply the graded Euler character of the space of local operators or, equivalently, a partition function of the theory on $S^2 \times S^1$ \cite{Kim:2009wb}.
There are several ways to compute this index: it is typically done using supersymmetric localization \cite{Kapustin:2011jm, Imamura:2011su} but, in the case of abelian gauge theories relevant to the present setting, it can also be computed directly from the space of states in a twisted version of the theory \cite{Zeng:2021zef}.
Either way, one arrives at the following result:
\be
\begin{aligned}
	I_{\textnormal{SCI}}[\CT_{N,k}](\eta;q) & = \Tr (-1)^{R} q^{J} \eta^S\\
	& = \sum\limits_{m_{ia} \in \Z} \oint \prod\limits_{i=1}^k \prod \limits_{a=1}^{2N-1} \frac{\textnormal{d} z_{ia}}{2\pi i z_{ia}} (-q^{-\scriptstyle{\frac{1}{2}}} \eta)^{(-1)^{a+1} i m_{ia}} z_{ia}^{\sum\limits_{j=1}^k\sum\limits_{b=1}^{2N-1} T_{ij}C_{ab} m_{jb}}\\
	& \qquad \qquad \times (-q^{\scriptstyle{\frac{1}{2}}} z_{ia}^{-1})^{\scriptstyle{\frac{1}{2}}(m_{ia} + |m_{ia}|)} \frac{(z_{ia}^{-1} q^{1+\scriptstyle{\frac{1}{2}}|m_{ia}|})_\infty}{(z_{ia} q^{\scriptstyle{\frac{1}{2}}|m_{ia}|})_\infty}
\end{aligned}
\ee
where we take the trace over the vector space of local operators, $R$ is the superconformal $R$-charge as determined by $F$-maximization \cite{Jafferis:2010un}, and $J = \frac{1}{2}R - J_3$ is the generator of twisted spin.
The integration contour is taken to be the product of unit circles $|z_{ia}|=1$.
As before, we use the notation $(x)_\infty = \prod_{n \geq 0} (1-x q^n)$.

We find that the above superconformal index starts as
\be
	I_{\textnormal{SCI}}[\CT_{N,k}](\eta;q) = 1 - q - \left(\eta + \eta^{-1}\right)q^{3/2} - 2 q^2 + \cdots\, .
\ee 
As discussed in \cite{Cordova:2016emh}, the existence of the $-(\eta + \eta^{-1}) q^{3/2}$ term in the index is a strong signal that the supersymmetry at the fixed point is enhanced to $\CN=4$: the only operators that can contribute this term either belong to additional supercurrent multiplets or are a pair of chiral primaries of $R$-charge $R=3$.
That the latter case does not happen for $N = k = 1$ was checked by the semi-classical analysis of \cite{Gang:2018huc} and for $N = 1$ and $k > 1$ in \cite{Gang:2021hrd}.
Furthermore, one can check that there exist two dressed monopole operators which can contribute to the $(\eta + \eta^{-1})q^{3/2}$ term in the superconformal index,
\be
    \left(\prod_{i=1}^k \phi_{i1}\phi_{i(2N-1)}\right)V_{\mathfrak{m}}\ ,\quad \mathfrak{m}_{ia} = -\delta_{i1} \quad (R=1,~S=-1,~j=1)\ ,
\ee
and
\be
    \psi^*_{k1}\phi_{k2}V_{\mathfrak{m}}\ , \quad \mathfrak{m}_{ia} = \left\{\begin{array}{cc} \delta_{1a}\delta_{1i}\ ,\quad k=1 \\ \delta_{1a}\delta_{ki}-\delta_{1a}\delta_{k-1,i}\ ,\quad k>1\end{array}\right.\quad (R=1,~S=1,~j=1)\ ,
\ee
whose quantum numbers coincide with those of the $\frac14$-BPS operators in the extra supercurrent multiplet, generalizing those proposed in \cite{Gang:2023rei}.\footnote{For $N>1$, we find that, semi-classically, there exist multiple operators with the same quantum numbers. From the superconformal index, we expect that they flow to the identical operators in the IR superconformal theory.}
We expect that the global symmetry $U(1)_R\times U(1)_S$ in the UV theory enhances to $SO(4) \simeq (SU(2)_H\times SU(2)_C)/\Z_2$ with the identification
\be
	R = J_3^H + J_3^C\ ,\quad S = J_3^C - J_3^H\ .
\ee
Additionally, we find that the superconformal indices of $\CT_{N,k}$ and $\CT_{k,N}$ coincide up to inverting $\eta$:
\be
	I_{\textnormal{SCI}}[\CT_{N,k}](\eta;q) \overset{?}{=} I_{\textnormal{SCI}}[\CT_{k,N}](\eta^{-1};q)
\ee
We have checked this claim to all orders in $q$ for $N = k = 1$ and to low orders in $q$ for all $N + k \leq 4$.

We can use the superconformal index to determine the Hilbert series counting local operators in the $A$ and $B$ twists as suitable specializations.
For example, in the $A$ twist we want to count local operators weighted by $(-1)^{R_A} q^{J_A}$, where $R_A$ and $J_A$ are the $R$-charge and twisted spin necessary for the $A$ twist: $R_A = 2 J_3^C = R + S$ and $J_A = J^H_3 - J_3 = J - \tfrac{1}{2}S$.
This can be obtained by specializing $\eta \to -q^{-\scriptstyle{\frac{1}{2}}}$ in the above superconformal index:
\be
	I_A[\CT_{N,k}](q) = \Tr (-1)^{R_A} q^{J_{A}} = \Tr (-1)^R q^{J} (-q^{-\scriptstyle{\frac{1}{2}}})^S = I_{\textnormal{SCI}}[\CT_{N,k}](-q^{-\scriptstyle{\frac{1}{2}}};q)\,.
\ee
Similarly, we can count local operators in the $B$ twist by specializing $\eta \to -q^{\scriptstyle{\frac{1}{2}}}$ in the above superconformal index:
\be
	I_B[\CT_{N,k}](q) = I_{\textnormal{SCI}}[\CT_{N,k}](-q^{\scriptstyle{\frac{1}{2}}};q)\,.
\ee
In both specializations we find that the superconformal index trivializes
\be
	I_A[\CT_{N,k}](q) = 1 = I_B[\CT_{N,k}](q)\ ,
\ee 
which is strong evidence that the IR SCFT is rank 0.

\subsection{Partition functions on Seifert manifolds and modular data}
Many other aspects of the topologically twisted theories $\CT^A_{N,k}$ and $\CT^B_{N,k}$ can be accessed using supersymmetric localization techniques.
We now describe one method for computing the partition functions of these TQFTs on Seifert manifolds and how to access modular properties of the state space on $T^2$.
As we describe below, this approach also gives access to the Grothendieck fusion rings of line operators in these twisted TQFTs.

The tool we use to compute these partition functions is to place the theory on $\R^2 \times S^1$ and study the low-energy dynamics of the resulting effective 2d $\CN = (2,2)$ theory on its Coulomb branch.
These dynamics are governed by an effective twisted superpotential $W$ and an effective dilaton coupling $\Omega$, both of which can be computed directly from the above UV $\CN=2$ Lagrangian theory \cite{Nekrasov:2009uh,Nekrasov:2014xaa,Closset:2017zgf}.
The effective twisted superpotential $W$ describes local aspects of the effective 2d theory on its Coulomb branch and takes the form
\be
\begin{aligned}
	W(z; \eta) =&~ \frac12 \sum_{I,J}\kappa_{IJ} \log z_{I} \log z_{J} + \log (-\eta) \sum_{I} (-1)^{a+1} i \log z_I + \sum_{I} \text{Li}_2(z_I)\,,
\end{aligned}
\ee
where $I = (i,a)$ and $J = (j,b)$ are multi-indices and Li${}_2(z)$ is the dilogarithm function.
Although the effective twisted superpotential suffers from branch-cut ambiguities, all physical quantities computed from it do not.
The dynamical variables $z_I$ are interpreted as (vacuum expectation values of complexified) holonomies around the $S^1$ for the dynamical $U(1)^{(2N-1)k}$ gauge fields and, similarly, the parameter $\eta$ is interpreted as the holonomy of a background $U(1)_S$ gauge field.
We note that the parameter $\eta$ is present to ensure the effective 2d theory remains fully massive but, just as we saw in the superconformal index, must be specialized as $\eta \to -1$ in order to reproduce the $A$ and $B$ twists.
All quantities should be computed for generic $\eta$ and then specialized $\eta \to -1$ at the end.

The supersymmetric vacua of the effective 2d theory can be determined by way of the Bethe ansatz \cite{Nekrasov:2009uh, Nekrasov:2009ui}.
They are given by solutions $z^* = z^*(\eta)$ to the Bethe equations
\be
	P_{I} := \exp\left[z_I \frac{\partial W}{\partial z_{I}}\right]=1\ ,~~\text{for all } I \,.
\ee
Explicitly, these equations read
\be
\label{eq:Bethe}
	\prod_J z_J^{\kappa_{IJ}} = (-\eta)^{(-1)^a i} (1-z_I)\,.
\ee
We find $\binom{N}{k}$ Bethe vacua for generic $\eta$.
From the perspective of the 3d TQFTs, these Bethe vacua can be interpreted as states on a torus $T^2$: for any $\CP(z)$, its value at a given Bethe vacua $z^*$ is interpreted as the expectation value of the corresponding line operator $\CL$ in this state.

The effective dilaton $\Omega$ describes how to couple the effective 2d theory to gravity and thus how to place it on a general Riemann surface.
This coupling is sensitive to which $R$-symmetry is used in the reduction, and in order to realize partition functions in the $A$ and $B$ twists we must use the corresponding $R$-symmetry.
When we use the $R$-symmetry generated by $R_\nu = R + \nu S$, the dilaton coupling takes the form
\be
	\Omega(z; \eta, \nu) = \sum_{I}\log(1-z_I) + (\nu-1) \sum_{I} (-1)^{a+1} i \log z_I\,.
\ee
From the effective dilaton and effective twisted superpotential, we define the handle-gluing and fibering operators
\be
\begin{aligned}
	\CH(z; \eta, \nu) =&~ \exp \left[\Omega(z;\eta, \nu)\right] \det_{I, J} \bigg(\frac{\partial^2 W}{\partial \log z_I \partial \log z_J}\bigg)\ ,\\
	\CF(z; \eta) =&~ \exp \left[W - \sum_I z_{I} \log z_I \frac{\partial W}{\partial z_I}- \eta \log \eta \frac{\partial W}{\partial \eta}\right]\,.
\end{aligned}
\ee
With these ingredients in hand, the partition function on a degree $p$ circle bundle over a genus $g$ Riemann surface takes the form
\be
\label{partition function}
	Z^{A}_{g,p} = \sum_{z^*} \CH^{A}(z^*)^{g-1} \CF^{A}(z^*)^p \qquad  Z^{B}_{g,p} = \sum_{z^*} \CH^{B}(z^*)^{g-1} \CF^{B}(z^*)^p\,,
\ee
where we set
\be
	\CH^{A}(z^*) = \lim\limits_{\eta \to -1} \CH(z^*; \eta, - 1) \qquad \CF^{A}(z^*) = \lim\limits_{\eta \to -1} \CF(z^*; \eta, - 1)\,.
\ee
and
\be
	\CH^{B}(z^*) = \lim\limits_{\eta \to -1} \CH(z^*; \eta, 1) \qquad \CF^{B}(z^*) = \lim\limits_{\eta \to -1} \CF(z^*; \eta, 1)\,.
\ee

We can extract modular data for the $A$- and $B$-twisted TQFTs from the form of these partition functions.
In particular, we identify the values of the (inverse of the) handle-gluing operator at the Bethe vacua $\{\CH^{B/A}(z^*)^{-1}\}$ with the first row of the $S$-matrix $\{(S^{B/A}_{0\mu})^{2}\}$ and the values of the fibering operator $\{\CF^{B/A}(z^*)\}$ with the (inverses of the) diagonal components of the $T$-matrix $\{(T^{B/A}_{\mu\mu})^{-1}\}$, up to an overall phase factor.%
\footnote{The overall phase of the fibering operator $\CF$ depends on contact terms for the background fields that we have not kept track of.} %
To extract other components of the $S$-matrix, we first note that there are two distinguished Bethe vacua $z^*_{0, \pm}$ corresponding to the trivial line operators in the $A$ and $B$ twists.
These distinguished Bethe vacua are constrained \cite{Cho:2020ljj} by the partition function of the $A$- and $B$-twisted theories on $S^3$ as%
\footnote{This condition need not have a unique solution as the handle-gluing operator can take the same value on different Bethe vacua, a phenomenon realized by \emph{e.g.} $\CT_{2,2}$.} %
\be
	\CH^{B/A}(z^*_{0,\pm})^{-1} = (S^{B/A}_{00})^2 = |Z^{B/A}_{0,1}|^2\,.
\ee
Once we have identified the Bethe vacuum corresponding to the trivial line operator $\CP^{B/A}_0 = 1$, we can identify the line operators $\CP^{B/A}_\mu(z)$ corresponding to the other Bethe vacua by requiring
\be
	\CP^{A}_{\mu}(z^*_{0,-}) = S^{A}_{0 \mu}/S^{A}_{00} \hspace{1cm} \CP^{B}_{\mu}(z^*_{0,+}) = S^{B}_{0 \mu}/S^{B}_{00}\,.
\ee
The remaining components of the $S$-matrix appear when evaluating these line operators on a general Bethe vacua
\be
\label{eq:BetheSmatrix}
	S^A_{\mu \nu} = \CP^{A}_{\mu}(z^*_{\nu, -}) S^A_{0\nu} \hspace{1cm} S^B_{\mu \nu} = \CP^{B}_{\mu}(z^*_{\nu, +}) S^A_{0\nu}\,.
\ee
This identity, together with the $PSL(2,\Z)$ relations and non-negativity of the fusion rules, determines the $S$-matrix up to an overall sign and the $T$-matrix up to an overall $3$rd root of unity.

\subsection{Bethe vacua and fusion rings}
The Bethe vacua themselves contain information about the fusion rules of line operators in the topologically twisted theories $\CT^A_{N,k}$ and $\CT^B_{N,k}$.
More precisely, the Grothendieck rings of the categories of line operators in these twisted theories are expected to coincide with the $\eta \to -1$ limit of the ring of Bethe vacua
\be
	\CR := \lim\limits_{\eta \to -1} \C[z_I^{\pm1}]\big/ \CI_\eta\,,
\ee
where $\CI_\eta$ is the ideal generated by the $P_I(z;\eta)$.
In particular, the polynomials $\CP^{A}_\mu(z)$ identified above reproduce the fusion rules of the corresponding line operators, \emph{i.e.} they satisfy
\be
\label{eq:Bethefusion}
	\CP^A_{\mu}(z^*) \CP^A_{\nu}(z^*) = \sum_{\lambda} N_{\mu \nu}^\lambda \CP^A_{\lambda}(z^*)
\ee
for all Bethe vacua $z^*$, where $N^\lambda_{\mu \nu}$ are the fusion coefficients of the simple line operators $\CL^A_\mu$.
The same is true for the $\CP^B_\mu$, albeit with (possibly) different fusion coefficients.
In view of Eq. \eqref{eq:BetheSmatrix}, this aspect of the Bethe vacua can be viewed as a consequence of the Verlinde formula, but is noteworthy because the ring $\CR$ does not have enough information to reproduce modular data by itself.

For the theories with $N = 1$ studied in \cite{Gang:2023rei} and \cite{Ferrari:2023fez}, these polynomials take the form
\be
\label{eq:N=1lines}
N = 1: \qquad \begin{aligned}
	\CP^A_{\mu} & = \prod\limits_{i=1}^{k} z_{i}{}^{-\min(\mu, i)}\\
	\CP^B_{\mu} & = (-1)^\mu \prod\limits_{i=0}^{\mu-1} z_{k-i}{}^{\mu-i}\\
\end{aligned}
\ee
for $\mu = 0, \dots, k$.
The factor of $(-1)^\mu$ corresponds to including an additional $R$-symmetry Wilson line.
For $k = 1$, the Bethe equations are still fairly simple and we find the polynomials
\be
\label{eq:k=1lines}
k = 1: \qquad \begin{aligned}
	\CP^A_{\mu} & = \prod\limits_{\alpha=0}^{\mu-1}(z_{2\alpha} z_{2\alpha+1}^{-1})\\
	\CP^B_{\mu} & = (-1)^\mu \prod\limits_{\alpha=1}^{\mu}(z_{2\alpha-1} z_{2\alpha}^{-1})\\
\end{aligned}
\ee
for $\mu = 0, \dots, N$ and we set $z_0 = z_{2N} = 1$.

We were unable to find closed-form expressions for these polynomials for general $N$ and $k$.
In the first non-edge case, \emph{i.e.} $N = k = 2$, we find the following line operators in the $A$ twist
\be
\begin{aligned}
	\CP^A_{(1,0)} = z_{13}^{-1}z_{23}^{-1} \qquad \CP^A_{(1,1)} = z_{13}^{-1} z_{23}^{-2} \qquad \CP^A_{(2,0)} = z_{11}^{-1} z_{21}^{-2} z_{12} z^{2}_{22} z_{13}^{-1} z_{23}^{-2}\big(z_{21} z_{23} + z_{22}\big)\\
	\CP^A_{(2,1)} = z_{11}^{-1} z_{21}^{-2} z_{12} z^{2}_{22} z_{13}^{-1} z_{23}^{-2}\big(z_{21} + z_{22} +  z_{23}\big) \qquad \CP^A_{(2,2)} = z_{11}^{-1} z_{12}^{-2} z_{21} z_{22}^{2} z_{31}^{-1} z_{32}^{-2} \hspace{0.5cm}\\
\end{aligned}
\ee
and the following line operators in the $B$ twist
\be
\begin{aligned}
	\CP^B_{(1,0)} = -(z_{21} + z_{22} + z_{23}) \qquad \CP^B_{(1,1)} = z_{21} z_{23} + z_{22} \qquad \CP^B_{(2,0)} = z_{11} z_{12}^2 z_{21}^{-1} z_{22}^{-2}\\
	\CP^B_{(2,1)} = - z_{11} z_{12}^2 z_{21}^{-1} z_{22}^{-2} z_{32} \qquad \CP^B_{(2,2)} = z_{11} z_{12}^2 z_{21}^{-1} z_{22}^{-2} z_{31} z_{32}^2 \hspace{2cm}\\
\end{aligned}
\ee
which, together with the trivial line operator $\CP^B_{(0,0)} = 1 = \CP^A_{(0,0)}$, realize the $6 = \binom{4}{2}$ Bethe vacua.
It is worth noting that, unlike the cases with $N = 1$ or $k = 1$ mentioned above, these polynomials are not all monomials, whence the corresponding line operators will not simply be Wilson lines.

We note that in all of these cases the polynomials arising in the $B$ twist are obtained from those in the $A$ twist by multiplying by an overall factor
\be
\label{eq:ABrelation}
	\CP^A_{\wt{\mu}} = (-1)^{Nk - |\mu|} \frac{\CP^B_\mu}{\CP^B_{\textnormal{top}}} \qquad \CP^B_{\textnormal{top}} = (-1)^{Nk}\prod_{i,a} z_{ia}^{(-1)^{a+1}i}\,,
\ee
which itself represents one of the $B$-twist line operators.

\subsection{3d mirror of $\CT_{N,k}$}
\label{sec:3dmirror}
The above analysis of $\CT_{N,k}^A$ and $\CT_{N,k}^B$ suggests a novel instance of 3d mirror symmetry involving these rank $0$ SCFTs. As mentioned above, we find that the superconformal index of $\CT_{N,k}$ agrees with that of $\CT_{k,N}$ after inverting $\eta$:
\be
	I_{\textnormal{SCI}}[\CT_{N,k}](\eta;q) = I_{\textnormal{SCI}}[\CT_{k,N}](\eta^{-1};q)\,.
\ee
Moreover, we find that the $A$-twist $S$-matrix of $\CT_{N,k}$ agrees with the $B$-twist $S$-matrix of $\CT_{k,N}$ (up to an overall sign that isn't fixed by the above analysis) and, correspondingly, the resulting fusion rings agree with one another.
Unfortunately, this agreement doesn't persist for the $T$-matrix: we find that the $A$-twist $T$-matrix of $\CT_{N,k}$ is the inverse of the $B$-twist $T$-matrix of $\CT_{k,N}$ (up to an overall $3$rd root of unity that isn't fixed by the above analysis).
This observation suggests that these theories differ by a parity transformation.
In fact, this parity reversal can be realized at the level of untwisted rank 0 SCFTs, roughly by negating the Chern-Simons levels of the UV $\CN=2$ theory.

We consider a theory $\ol{\CT}_{N,k}$ realized as above by deforming a 3d $\CN=2$ theory of $k(2N-1)$ copies of $U(1)$ gauge theory with a charge $1$ chiral multiplet, coupled to one another with the mixed Chern-Simons levels
\be
	\ol{\kappa} = \id - \kappa\,.
\ee
For example, $\ol{\CT}_{1,1}$ is simply $U(1)_{-3/2}$ plus a chiral of charge $1$, which was predicted by \cite{Gang:2018huc} to exhibit an enhancement to $\CN=4$ supersymmetry.
The superconformal index of this theory is exactly the same as $\CT_{N,k}$, but with $m_{ia} \to -m_{ia}$.
\be
\begin{aligned}
	I_{\textnormal{SCI}}[\ol{\CT}_{N,k}](\eta;q) & = \sum\limits_{m_{ia} \in \Z} \oint \prod\limits_{i=1}^k \prod \limits_{a=1}^{2N-1} \frac{\textnormal{d} z_{ia}}{2\pi i z_{ia}} \bigg((-q^{-\scriptstyle{\frac{1}{2}}}) (-q^{-\scriptstyle{\frac{1}{2}}}\eta)^{(-1)^{a} i}\bigg)^{m_{ia}} z_{ia}{}^{m_{ia}-\sum\limits_{j=1}^k\sum\limits_{b=1}^{2N-1} T_{ij}C_{ab} m_{jb}}\\
	& \qquad \qquad \times (-q^{\scriptstyle{\frac{1}{2}}} z_{ia}^{-1})^{\scriptstyle{\frac{1}{2}}(m_{ia} + |m_{ia}|)} \frac{(z_{ia}^{-1} q^{1+\scriptstyle{\frac{1}{2}}|m_{ia}|})_\infty}{(z_{ia} q^{\scriptstyle{\frac{1}{2}}|m_{ia}|})_\infty}\\
\end{aligned}
\ee
We thus find
\be
	I_{\textnormal{SCI}}[\ol{\CT}_{k,N}](\eta;q) = I_{\textnormal{SCI}}[\CT_{k,N}](\eta;q) = I_{\textnormal{SCI}}[\CT_{N,k}](\eta^{-1};q)\,.
\ee
From this expression for the superconformal index, we find that the generators of superconformal $R$-charge and $S$-charge are given by
\be
	R = R' - \sum_{a=1}^{2N-1}\big(1-(-1)^{a+1}i\big)\ol{M}_{ia} \qquad S = \sum_{a=1}^{2N-1}\sum_{i=1}^k (-1)^a i \ol{M}_{ia}
\ee
where $\ol{M}_{ia}$ are the generators of the topological flavor symmetries of the UV gauge theory and $R'$ is the UV $R$-charge that gives the chiral multiplet scalars weight $0$.

Using this superconformal $R$-charge and $S$-charge, we find an effective twisted superpotential given by%
\footnote{The factors of $-1$ in the effective twisted superpotential accounts for specializing the fugacities $\ol{t}_{ia}$ for the topological flavor symmetry as $\ol{t}_{ia} \to (-1)^{1+(-1)^a i} \eta^{(-1)^a i} = - (-\eta)^{(-1)^a i}$.} %
\be
	\ol{W} = \tfrac{1}{2} \sum_{I,J} \ol{\kappa}_{IJ} \log z_I \log z_J  + \sum_I \log(-(-\eta)^{(-1)^{a} i}) \log z_I + \textnormal{Li}_2(z_I)
\ee
and the effective dilaton coupling using $R_\nu = R + \nu S$ is given by
\be
	\ol{\Omega} = \sum_I \log(1-z_I)-(1-(1-\nu)(-1)^{a+1}i) \log z_I\,.
\ee
Performing the above Bethe analysis, we are lead to the following equivalence of TQFTs
\be
	\CT^A_{N,k}	\simeq \overline{\CT}^{B}_{k,N} \qquad \CT^B_{N,k}	\simeq \overline{\CT}^{A}_{k,N}\,.
\ee
The line operators in $\CT_{N,k}$ and in $\ol{\CT}_{N,k}$ are related as
\be
	\ol{\CP}^A_\mu(z) = \CP^A_\mu(z^{-1}) \qquad \ol{\CP}^B_\mu(z) = \CP^B_\mu(z^{-1})\,.
\ee
Together with the above superconformal index, it is natural to expect that these equivalences of TQFTs actually arise from an instance of 3d mirror symmetry for rank 0 SCFTs:
\be
	\textnormal{3d mirror symmetry:} \quad \CT_{N,k} \longleftrightarrow \overline{\CT}_{k,N} = (\CT_{N,k})^\vee
\ee

\section{\texorpdfstring{$\CT^B_{N,k}$}{TNkB} and \texorpdfstring{$L_k(\fosp_{1|2N})$}{Lkosp12N}}
\label{sec:osp}
The $B$ twist of $\CT_{k} = \CT_{1,k}$ was argued to have a holomorphic boundary condition furnishing the simple affine VOA $L_k(\fosp_{1|2})$ in \cite{Ferrari:2023fez}.
In this section we argue for a natural generalization to the $B$-twisted theory $\CT^B_{N,k}$: it admits a holomorphic boundary condition Dir${}_{N,k}$ furnishing the simple affine VOA $L_k(\fosp_{1|2N})$.

As in \cite{Ferrari:2023fez}, we start with a boundary condition of the UV $\CN=2$ theory described in Section \ref{sec:TNk} that we expect will flow to a boundary condition for $\CT_{N,k}$ that is deformable (in the sense of \cite{Costello:2018fnz, Brunner:2021tfl}) to its $B$ twist.
The form of this UV boundary condition is highly constrained by the superpotential used to deform to $\CT_{N,k}$ and by the form of the supercharge used in the $B$ twist; see Section 3 of \cite{Ferrari:2023fez} for more details.
The UV boundary condition of interest imposes $\CN=(0,2)$ Dirichlet boundary conditions on both the vector multiplets and the chiral multiplets, with most of the latter given \emph{generic} Dirichlet boundary conditions --- the chiral multiplet scalars $\phi_{ia}$ are set to generic, non-zero values if $i < k$ or if $a$ is even, and are otherwise required to vanish, \emph{i.e.} we set $\phi_{k(2\alpha-1)}| = 0$ for $\alpha = 1, \dots , N$ and $\phi_{ia}| = c_{ia} \neq 0$ otherwise.

We now present several pieces of evidence that the algebra of local operators on this boundary condition realizes the simple affine VOA $L_k(\fosp_{1|2N})$ upon deformation to $\CT^B_{N,k}$.

\subsection{Half-indices and modular Nahm sums}
For an $\CN=(0,2)$ boundary condition $\CB$ of a 3d $\CN=2$ theory, the half-index \cite{Gadde:2013sca, Yoshida:2014ssa, Dimofte:2017tpi} is a signed count of local operators on $\CB$
\be
	I\!\!I_\CB(x_\alpha; q) = \Tr(-1)^R q^J \prod\limits_{\alpha} x_\alpha^{N_\alpha}
\ee
where $R$ is the generator of the $\CN=2$ $R$-symmetry, $J = \frac{1}{2} R - J_3$ is the generator of twisted spin, and $N_\alpha$ are commuting flavor symmetries of the boundary algebra.

The half-index $I\!\!I_{N,k}^B$ counting local operators on $\textnormal{Dir}_{N,k}$ in the $B$-twisted theory $\CT^B_{N,k}$ can be obtained as a suitable specialization of the half-index $\wt{I\!\!I}_{N,k}$ counting local operators on the Dirichlet boundary condition of the UV theory that sets all of the chiral multiplet scalars to zero, \emph{cf.} \cite[Section 3.2.2]{Dimofte:2017tpi}.
\be
\label{eq:HTBhalfindex}
\begin{aligned}
	\wt{I\!\!I}_{N,k}(z_{ia}, t_{ia};q) & = \sum_{m_{ia} \in \mathbb{Z}} q^{\scriptstyle{\frac{1}{2}} \sum\limits_{i,j =1}^{k}\sum\limits_{a,b=1}^{2N-1} T_{ij} C_{ab}m_{ia}m_{jb}}\\
	& \qquad \times \prod\limits_{i=1}^k \prod\limits_{a=1}^{2N-1} ((-q^{-\scriptstyle{\frac{1}{2}}})^{(-1)^{a+1} i}t_{ia})^{m_{ia}} z_{ia}^{\sum\limits_{j=1}^k\sum\limits_{b=1}^{2N-1} T_{ij} C_{ab} m_{jb}} \frac{(q^{1-m_{ia}} z_{ia}^{-1})_\infty}{(q)_\infty}
\end{aligned}
\ee
The integers $m_{ia}$ correspond to magnetic charges of boundary local operators and the fugacities/Jacobi variables $t_{ia}$ measure charges for the topological flavor symmetry dual to $U(1)_{ia}$ generated by $M_{ia}$, \emph{i.e.} the magnetic charge $m_{ia}$.
The $z_{ia}$ are fugacities/Jacobi variables for boundary $U(1)_{ia}$ gauge transformations.
As with the superconformal indices in Section \ref{sec:TNk}, the $R$-charge used in this expression for the half-index is the superconformal $R$-charge.

The superpotential used to deform the UV theory to $\CT_{N,k}$ requires the boundary values for most of the $\phi_{ia}$ to be non-zero.
A non-zero boundary value for $\phi_{ia}$ breaks the boundary $U(1)_{ia}$ symmetry and we must correspondingly specialize $z_{ia} = 1$; the UV boundary condition described above only preserves $U(1)_{k(2\alpha-1)}$ for $\alpha = 1, \dots, N$.
The superpotential also breaks nearly all of the topological flavor symmetries, leaving only $S = \sum (-1)^{a+1} i M_{ia}$.
Correspondingly, if we wish to count local operators in $\CT_{N,k}$, we should further specialize $t_{ia} = \eta^{(-1)^{a+1} i}$.
Finally, passing to the $B$ twist breaks $U(1)_S$ and requires a change in twisted spin and $R$-charge: we must finally specialize $\eta = - q^{\scriptstyle{\frac{1}{2}}}$, as we did in Section \ref{sec:TNk}.
Putting this together, the half-index $I\!\!I_{N,k}^B$ counting local operators on Dir${}_{N,k}$ can be written as the following specialization of $\wt{I\!\!I}_{N,k}(z_{ia}, t_{ia};q)$:
\be
	I\!\!I^B_{N,k}(x_\alpha;q) = \wt{I\!\!I}_{N,k}(z_{ia}, t_{ia};q)\bigg\vert_{(\star)}
\ee
where $(\star)$ denotes specializing the Jacobi variables as $t_{ia} = (-q^{\scriptstyle{\frac{1}{2}}})^{(-1)^{a+1}i}$ for all $i,a$ and $z_{ia}=1$ if $i < k$ or if $a$ is even; we denote $x_\alpha = z_{k (2\alpha -1)}$ for $\alpha = 1, \dots, N$.
Explicitly, we have
\be
	I\!\!I^B_{N,k}(x_\alpha;q) = \sum\limits_{m_{ia}}{}' q^{\scriptstyle{\frac{1}{2}} \sum\limits_{i,j =1}^{k}\sum\limits_{a,b=1}^{2N-1} T_{ij} C_{ab}m_{ia}m_{jb}}\frac{\prod\limits_{\alpha=1}^N (q^{1-m_{k(2\alpha-1)}} x_{\alpha}^{-1})_\infty x_{\alpha}^{\sum\limits_{j=1}^{k} j (2m_{j(2\alpha-1)}-m_{j(2\alpha)}-m_{j(2\alpha-2)})}}{(q)_\infty^N \left(\prod\limits_{i=1}^{k-1} \prod\limits_{a=1}^{2N-1}(q)_{-m_{ia}}\right) \left(\prod\limits_{\beta =1}^{N-1}(q)_{-m_{k (2\beta)}}\right)}
\ee
where $\sum'$ refers to a summation with $m_{ia}$ a non-positive integer if $i < k$ or if $a$ is even, and any integer otherwise; we set $m_{i0} = m_{i(2N)} = 0$ for all $i$.
It was shown in \cite{Ferrari:2023fez} that this expression for $N = k = 1$ reproduces the (normalized) vacuum character of $L_1(\fosp_{1|2})$ and its equality with the (normalized) vacuum character of $L_k(\fosp_{1|2})$ was checked when $N = 1$ and $k \leq 3$ to order $q^{10}$.
More generally, we expect that this half-index matches the (normalized) vacuum character of $L_k(\fosp_{1|2N})$:
\be
	I\!\!I^B_{N,k}(x_\alpha;q) \overset{?}{=} \ol{\chi}[L_k(\fosp_{1|2N})](x_\alpha;q)\,.
\ee
We have checked this expectation to order at least $q^5$ for all $N$, $k$ with $N+k \leq 4$.

Further evidence for the appearance of $L_k(\fosp_{1|2N})$ comes from specializing the remaining Jacobi variables $x_\alpha \to 1$. The above half-index naturally leads to an expression for the vacuum character as a Nahm sum based on the matrix $\kappa$:
\be
	I\!\!I^B_{N,k}(1;q) = \sum_{n_{ia} \in \mathbb{Z}_{\geq 0}} \frac{q^{\scriptstyle{\frac{1}{2}} \sum\limits_{i,j =1}^{k}\sum\limits_{a,b=1}^{2N-1} T_{ij} C_{ab}n_{ia}n_{jb}}}{\prod\limits_{i=1}^k\prod\limits_{a=1}^{2N-1} (q)_{n_{ia}}}
\ee
This expression for the (normalized, unrefined) vacuum character of $L_k(\fosp_{1|2N})$ as a Nahm sum of type $(A_{2N-1}, T_k)$ is not new: it also appears in work of Warnaar and Zudilin, \emph{cf.} Conjecture 1.1 and Theorem 1.2 of \cite{Warnaar2012} where this identity is conjectured and proven when $k = 1$.
For $N = 1$, these Nahm sums are related to Rogers-Ramanujan and Andrews-Gordon identities \cite{Andrews_1984} and to characters of the minimal models $M(2,2k+3)$.
For $k = 1$, they are known to be modular for all $N$ \cite{Bringmann2014}.
More generally, Nahm sums based on a bilinear form $\kappa$ of the form $C(Y)^{-1} \otimes C(X)$ for $X,Y$ Dynkin diagrams of type ADET are expected to enjoy modular properties; see \emph{e.g.} \cite{Gliozzi:1994cs, Nahm:2004ch} and references therein.
The effective central charge for the resulting vector-valued modular function is conjecturally determined by the Lie algebras $X,Y$ as
\be
	c_{X,Y} = \frac{\textnormal{rank}(X) \textnormal{rank}(Y) h^\vee_X}{h^\vee_X + h^\vee_Y}
\ee
where $h^\vee_X$, $h^\vee_Y$ are the dual Coxeter numbers of $X$, $Y$.
When we take $X = A_{2N-1}$ and $Y = T_k$ this is precisely the Sugawara central charge of $L_k(\fosp_{1|2N})$
\be
	c^B_{N,k} = c_{A_{2N-1}, T_k} = \frac{(2N-1) k (2N)}{2 N + 1 + 2 k} = \frac{k \, \textnormal{sdim}(\fosp_{1|2N})}{k + h^\vee_{\fosp_{1|2N}}}.
\ee
In recent work \cite{Creutzig:2025tgh} by the first two authors, it is shown that this specialization indeed reproduces the (normalized) vacuum character:
\be
    I\!\!I^B_{N,k}(1;q) = \ol{\chi}[L_k(\fosp_{1|2N})](1;q)\,.
\ee

\subsection{Nahm's equations, Bethe vacua, and $L_k(\fosp_{1|2N})$ modules}

With the appearance of $L_k(\fosp_{1|2N})$ at the boundary of $\CT^B_{N,k}$, it is natural to expect that its modules can be realized by bulk line operators ending on Dir${}_{N,k}$.
This expectation is again supported by computing half-indices, which reproduce supercharacters of $L_k(\fosp_{1|2N})$ modules.

The half-index counting local operators at the junction of a line operator $\CL$ and the boundary condition Dir${}_{N,k}$ can again be written as a specialization, \emph{cf.} Eq. (3.43) of \cite{Dimofte:2017tpi}.
When we further specialize the remaining Jacobi variables $x_\alpha = 1$, we generally find linear combinations of Nahm sums, all based on the same bilinear form $\kappa$. There are deep connections between the modularity of these Nahm sums and number-theoretic properties of solutions to certain polynomial equations called Nahm's equations \cite{Nahm:2004ch, Zagier:2007knq}:
\be
\label{eq:Nahm}
	\prod_{J} z_{J}^{\kappa_{IJ}} = 1 - z_I
\ee
for each (multi-)index $I = (i,a)$.%
\footnote{We note that the Nahm's conjecture, relating modular Nahm sums to torsion elements in the Bloch group of the field of rational number adjoint solutions to these equations, has been proven in some cases \cite{Calegari:2017itq} but is provably false \cite{VlasenkoZwegers} as originally formulated.} %
These are precisely the same equations governing 3d Bethe vacua of the theory $\CT_{N,k}$, \emph{cf.} Eq. \eqref{eq:Bethe}, and whose solutions (after a suitable regularization) are identified with simple line operators $\CL^B_\mu$ of $\CT^B_{N,k}$.
The analysis of these equations in Section \ref{sec:TNk} indicates that there are $\binom{N}{k}$ such vacua, precisely matching the number of simple modules of $L_k(\fosp_{1|2N})$ \cite{Creutzig:2022riy}, and for each a corresponding generator $\CP^B_{\mu}(z_{ia})$ of the ring of Bethe vacua realizing the Grothendieck ring of the category of lines.
In the explicit examples, we have chosen the labels of these line operators to agree with those of the labeling of simple $L_k(\fosp_{1|2N})$ modules given in \cite{Creutzig:2022riy}.%
\footnote{In the cases with $k = 1$ we use $\mu$ to denote the number of $1$'s in the label.} %

As mentioned above, we can compute the half-index counting local operators at the end of each of these generators as a suitable specialization:
\be
	I\!\!I^B_{N,k}[\CL^B_\mu](x_\alpha;q) = \wt{I\!\!I}_{N,k}[\CL^B_\mu](z_{ia}, t_{ia};q)\big\vert_{(\star)}
\ee
where
\be
\begin{aligned}
	\wt{I\!\!I}_{N,k}[\CL^B_\mu](z_{ia}, t_{ia};q) & = \sum_{m_{ia} \in \mathbb{Z}} q^{\scriptstyle{\frac{1}{2}} \sum\limits_{i,j =1}^{k}\sum\limits_{a,b=1}^{2N-1} T_{ij} C_{ab}m_{ia}m_{jb}} \CP^B_\mu(q^{-m_{ia}} z_{ia}^{-1})\\
	& \qquad \times \prod\limits_{i=1}^k \prod\limits_{a=1}^{2N-1} ((-q^{-\scriptstyle{\frac{1}{2}}})^{(-1)^{a+1} i}t_{ia})^{m_{ia}} z_{ia}^{\sum\limits_{j=1}^k\sum\limits_{b=1}^{2N-1} T_{ij} C_{ab} m_{jb}} \frac{(q^{1-m_{ia}} z_{ia}^{-1})_\infty}{(q)_\infty}
\end{aligned}\,.
\ee
For $N=1$, these expressions were conjectured to reproduce the characters of all simple modules of $L_k(\fosp_{1|2})$ in \cite{Ferrari:2023fez}, where it was proven to be true for $k = 1$ and checked to order $q^{10}$ for all $k \leq 3$.
More generally, we have checked that the $\CP^B_\mu$ presented in Section \ref{sec:TNk} for all $N$, $k$ with $N+k \leq 4$ indeed reproduce characters of $L_k(\fosp_{1|2N})$, at least to order $q^5$.

It is worth noting that the fully-specialized half-indices for the Wilson lines in the two edge cases ($N = 1$ and $k = 1$) are very closely related to the form of the Nahm sums appearing in the left hand side of Conjecture 1.1 and Theorem 1.2 of \cite{Warnaar2012}.
Based on our findings in Section \ref{sec:TNk} and those of \cite{Warnaar2012}, we expect that the line operator $\CP^{B}_{\textnormal{top}}$ corresponds to the $L_k(\fosp_{1|2N})$ module with label $\mu = (k, \dots, k)$.
At the level of supercharacters, this translates to the following identity.
\be
\begin{aligned}
	\ol{\chi}[L_k(\mu)](x_\alpha; q) & \overset{?}{=} (-1)^{Nk}\sum\limits_{m_{ia}}{}' q^{\scriptstyle{\frac{1}{2}} \sum\limits_{i,j=1}^{k}\sum\limits_{a,b=1}^{2N-1} T_{ij} C_{ab}m_{ia}m_{jb} + \sum\limits_{i=1}^k\sum\limits_{a=1}^{2N-1} (-1)^a i m_{ia}}\\
	& \qquad \times \frac{\prod\limits_{\alpha=1}^N (q^{1-m_{k(2\alpha-1)}} x_{\alpha}^{-1})_\infty x_{\alpha}^{-k + \sum\limits_{j=1}^{k} j (2m_{j(2\alpha-1)}-m_{j(2\alpha)}-m_{j(2\alpha-2)})}}{(q)_\infty^N \left(\prod\limits_{i=1}^{k-1} \prod\limits_{a=1}^{2N-1}(q)_{-m_{ia}}\right) \left(\prod\limits_{\beta =1}^{N-1}(q)_{-m_{k (2\beta)}}\right)}
\end{aligned}
\ee

To end this section, we turn to the comparison of fusion rules of $L_k(\fosp_{1|2N})$ and those computed by the Bethe analysis of Section \ref{sec:TNk}, at least in the cases where we have explicit predictions for the generators of the fusion ring.
For $N=1$, the fusion rules were originally described in \cite{Creutzig:2018zqd} by realizing $L_k(\fosp_{1|2})$ as an extension of $L_k(\fsl_2) \otimes M(k+2,2k+3)$, see \emph{e.g.} Eq. (5.14) of \cite{Ferrari:2023fez} for an explicit formula.
More generally, the fusion rules for $L_k(\fosp_{1|2N})$ are equivalent to those of (a certain subcategory of modules for) the principal $W$-algebra $W_\ell(\fsp_{2N})$, where the level $\ell$ satisfies
\be
    \frac{1}{k + N + 1} + \frac{1}{\ell + N + 1} = 2\,.
\ee
The fusion rules for this $W$-algebra were first determined by \cite{Frenkel:1992ju}, see also \cite{Creutzig:2019qje, Arakawa:2019ear}.
It is a straightforward, albeit tedious, task to verify that the representative polynomials to reproduce these fusion rules for simple cases.
We have not been able to verify that they satisfy the necessary fusion rules for all of the examples presented in Section \ref{sec:TNk}, but have checked it agrees for all $(N, k)$ satisfying $N + k \leq 4$.
As a illustration, consider $N = 1$,  $k = 2$, where we have the relations $z_1^2 z_2^4 = 1-z_2$ and $z_1^2 z_2^2 = 1 - z_1$.
If we write $w_1 := \CP^B_1 = -z_2$ and $w_2 = \CP^B_2 = z_1 z_2^2$, then these relations become
\be
    w_2^2 = 1 + w_1 \qquad w_1^{-2} w_2^2 = 1 - w_1^{-2} w_2\,.
\ee
Multiplying the second relation by $w_1^2$ and plugging in the first gives $w_1^2 = 1 + w_1 + w_2$.
We can perform similar manipulations to deduce $w_1 w_2 = w_1 + w_2$:
\be
    1 = w_1^{-1}(w_2^2 - 1) = w_1^{-1}(w_2 + 1)(w_2 - 1) = (w_1 - 1)(w_2 - 1)\,.
\ee
We conclude that in the ring of Bethe vacua we have
\be
    w_1^2 = 1 + w_1 + w_2 \qquad w_1 w_2 = w_1 + w_2 \qquad w_2^2 = 1 + w_1
\ee
which is precisely the fusion ring for $L_2(\fosp_{1|2})$.

\section{\texorpdfstring{$\CT^A_{N,k}$}{TNkA} and \texorpdfstring{$W^{\textnormal{min}}_{k-\scriptstyle{\frac{1}{2}}}(\fsp_{2N})$}{Wminsp}}
\label{sec:Wmin}
We now turn to the $A$-twisted theory $\CT^A_{N,k}$.
The work \cite{Gang:2023rei} argued that the $A$ twist of $\CT_k = \CT_{1,k}$ has a holomorphic boundary condition furnishing the Virasoro minimal model $M(2,2k+3)$.
In this section, we present evidence for a generalization to the family of theories $\CT_{N,k}$: the $A$-twisted theory $\CT_{N,k}^A$ has a holomorphic boundary condition furnishing the minimal $W$-algebra of $\fsp_{2N}$ at level $k - \frac{1}{2}$.

The boundary condition studied in \cite{Gang:2023rei} for the $A$ twist when $N = 1$ is very similar to, and indeed inspired, the boundary condition studied in \cite{Ferrari:2023fez} for the $B$ twists of these same theories.
With this in mind, it is natural to expect that this feature continues to hold more generally: the UV boundary condition we consider for the $A$ twist will only be a slight modification of the one used above for the $B$ twist.
In particular, we will simply deform the boundary condition on the scalar $\phi_{k1}$ from requiring it to vanish at the boundary to instead requiring it to acquire a non-vanishing, generic value.
We expect that this UV boundary condition will flow to a boundary condition of $\CT_{N,k}$ that is deformable to the $A$ twist.

\subsection{Half-indices and characters}

The half-index counting local operators on Dir${}_{N,k}'$ in the $A$ twist can also be obtained as a specialization of $\wt{I\!\!I}_{N,k}(z_{ia}, t_{ia};q)$.
The non-vanishing $\phi_{ia}|$ break the $U(1)_{ia}$ symmetry if $i < k$, if $a$ is even, or if $a = 1$; this will again force us to set most of the $z_{ia}$ to $1$.
The topological flavor symmetries are also fully broken after we pass to the $A$ twist and so we must specialize the $t_{ia}$, but we should not specialize them exactly the same way we did in the previous section --- we again specialize $t_{ia} = \eta^{(-1)^{a+1}i}$ to count local operators in $\CT_{N,k}$ but then must specialize $\eta = - q^{-\scriptstyle{\frac{1}{2}}}$ to count local operators in the $A$ twist, \emph{cf.} Section \ref{sec:TNk}.
We are thus lead to the following expression for the half-index:
\be
	I\!\!I^A_{N,k}(w_\nu; q) = \wt{I\!\!I}_{N,k}(z_{ia}, t_{ia};q)\bigg\vert_{(\star \star)}
\ee
where $(\star \star)$ denotes specializing $t_{ia} = (-q^{-\scriptstyle{\frac{1}{2}}})^{(-1)^{a+1} i}$ and specializing $z_{ia} = 1$ if $i < k$, if $a$ is even, or if $a = 1$.
We also denote $z_{k (2\nu+1)} = w_\nu$ for $\nu = 1, \dots, N-1$.
The half-index counting local operators on Dir${}'_{N,k}$ in the $A$ twist is thus given by
\be
\begin{aligned}
	I\!\!I^A_{N,k}(w_\nu;q) & = \sum\limits_{m_{ia}}{}'' q^{\scriptstyle{\frac{1}{2}} \sum\limits_{i,j =1}^{k}\sum\limits_{a,b=1}^{2N-1} T_{ij} C_{ab}m_{ia}m_{jb} + \sum\limits_{i=1}^k \sum\limits_{a=1}^{2N-1}(-1)^a i m_{ia}}\\
	& \qquad \times \frac{\prod\limits_{\nu=1}^{N-1} (q^{1-m_{k(2\nu+1)}} w_{\nu}^{-1})_\infty w_\nu^{\sum\limits_{j=1}^{k} j (2m_{j(2\nu+1)}-m_{j(2\nu+2)}-m_{j(2\nu)})}}{(q)_\infty^{N-1} \left(\prod\limits_{i=1}^{k-1} \prod\limits_{a=1}^{2N-1}(q)_{-m_{ia}}\right) \left((q)_{-m_{k 1}}\prod\limits_{\beta=1}^{N-1}(q)_{-m_{k (2\beta)}}\right)}
\end{aligned}
\ee
where $\sum''$ refers to a summation where $m_{ia}$ is a non-positive integer if $i < k$, if $a$ is even, or if $a = 1$, and any integer otherwise.

That these half-indices reproduce the (normalized) vacuum characters of $M(2,2k+3)$ for $N = 1$ was observed in \cite{Gang:2023rei} and relies on the fermionic sum representations of $M(2,2k+3)$ characters due to \cite{Feigin:1991wv, Nahm:1992sx, Kedem:1993ze, Berkovich:1994es, Nahm:1994vas}.
For $N > 1$, we find that the above half-index agrees with the (normalized) vacuum character of the minimal $W$-algebra, but with an unusual stress tensor obtained by spectral flow. Namely, we expect that sending $w_\nu \to q^{\scriptstyle{\frac{1}{2}}} w_\nu$ in the above expression reproduces the (normalized) vacuum character with its usual stress tensor:
\be
	I\!\!I^A_{N,k}(q^{\scriptstyle{\frac{1}{2}}} w_\nu;q) \overset{?}{=} \ol{\chi}[W^{\textnormal{min}}_{k-\scriptstyle{\frac{1}{2}}}(\fsp_{2N})](w_\nu;q)
\ee
which we have verified to order $q^4$ for values of $N$ and $k$ with $N+k \leq 4$.

Just like the $B$ twist studied in Section \ref{sec:osp}, further specializing $w_\nu = 1$ yields a Nahm sum.
Namely, we get (up to an overall sign) the Nahm sum coming from the line operator $\CP^B_{\textnormal{top}}$ that we expect to reproduce the character of the $L_k(\fosp_{1|2N})$ module $L_k(\mu)$ for $\mu = (k, \dots, k)$.
This is precisely the second Nahm sum appearing in Conjecture 1.1 and Theorem 1.2 of \cite{Warnaar2012}.
It was conjectured, and proven for $k =1$, in \cite{Bringmann2014} that these sums are modular with modular anomaly $q^{-c^A_{N,k}/24}$ for
\be
	c^A_{N,k} = -\frac{k N (1+ 6 k + 4 N)}{k + N + \frac{1}{2}} = \bigg(\frac{(k-\frac{1}{2})N(2N+1)}{k + N + \frac{1}{2}} + N - 6 k\bigg) - 6 k(N-1)\,.
\ee
The expression in the parenthesis of the last equality is the central charge of $W^{\textnormal{min}}_{k-\scriptstyle{\frac{1}{2}}}(\fsp_{2N})$ \cite{Kac2003} and the remaining $-6k(N-1)$ accounts for the change in conformal vector due to spectral flow.
That the usual conformal vector is not appropriate for the $A$ twist can also be seen from the Bethe analysis of Section \ref{sec:TNk}: the values of $T_{00}$ match $e^{-i \pi c^A_{N,k}/12}$, not the usual central charge.

\subsection{Line operators and $W^{\textnormal{min}}_{k-\scriptstyle{\frac{1}{2}}}(\fsp_{2N})$ modules}

We can also recover characters of $W^{\textnormal{min}}_{k-\scriptstyle{\frac{1}{2}}}(\fsp_{2N})$ modules as half-indices counting local operators at the end of bulk line operators.
Much of the discussion in Section \ref{sec:osp} carries over to this case, so we will be brief.
The salient points are that we expect that the line operators identified in Section \ref{sec:TNk} lead to simple modules for this $W$-algebra and that the characters of those modules are reproduced by the corresponding half-indices.
Explicitly, these half-indices take the form
\be
	I\!\!I^A_{N,k}[\CL^A_\mu](w_\nu;q) = \wt{I\!\!I}_{N,k}[\CL^A_\mu](z_{ia}, t_{ia};q)\bigg\vert_{(\star \star)}
\ee
where
\be
\begin{aligned}
	\wt{I\!\!I}_{N,k}[\CL^A_\mu] & = \sum_{m_{ia} \in \mathbb{Z}} q^{\scriptstyle{\frac{1}{2}} \sum\limits_{i,j =1}^{k}\sum\limits_{a,b=1}^{2N-1} T_{ij} C_{ab}m_{ia}m_{jb}} \CP^A_\mu(q^{-m_{ia}} z_{ia}^{-1})\\
	& \qquad \times \prod\limits_{i=1}^k \prod\limits_{a=1}^{2N-1} ((-q^{-\scriptstyle{\frac{1}{2}}})^{(-1)^{a+1} i}t_{ia})^{m_{ia}} z_{ia}^{\sum\limits_{j=1}^k\sum\limits_{b=1}^{2N-1} T_{ij} C_{ab} m_{jb}} \frac{(q^{1-m_{ia}} z_{ia}^{-1})_\infty}{(q)_\infty}
\end{aligned}\,.
\ee
That these half-indices reproduce the (normalized) characters of all $M(2,2k+3)$ modules for $N = 1$ was observed in \cite{Gang:2023rei} and again relies on the aforementioned fermionic sum representations of $M(2,2k+3)$ characters.

It was noticed in \cite{Ferrari:2023fez} that for $N = 1$ these half-indices obey
\be
	I\!\!I^A_{1,k}[\CL^A_\mu](q) = (-1)^{k-\mu}I\!\!I^B_{1,k}[\CL^B_{k-\mu}](1;q)
\ee
thereby relating $M(2,2k+3)$ characters and specializations of $L_k(\fosp_{1|2})$ characters.
Based on the observation made at the end of Section \ref{sec:TNk}, in particular the relation between the $A$ and $B$ twist line operators in Eq. \eqref{eq:ABrelation}, we expect this feature to continue.
Namely, specializations of normalized $L_k(\fosp_{1|2N})$ characters should agree (up to an overall sign) with specializations of $W^{\textnormal{min}}_{k-\scriptstyle{\frac{1}{2}}}(\fsp_{2N})$ modules:
\be
	(-1)^{|\mu|}\ol{\chi}[L_k(\mu)](1;q) \overset{?}{=} \ol{\chi}[W^{\textnormal{min}}_{k-\scriptstyle{\frac{1}{2}}}(\wt{\mu})](q^{-\scriptstyle{\frac{1}{2}}};q)\,.
\ee

Unlike the $B$-twist, we cannot check that the rings of Bethe vacua described in Section \ref{sec:TNk} match the expectations for the fusion rings of the minimal $W$-algebras $W^{\rm min}_{k-\scriptstyle{\frac{1}{2}}}(\fsp_{2N})$ as the latter are not known, except when $N = 1$ where these $W$-algebras are simply the Virasoro minimal models $M(2, 2k+3)$.
We can still determine the ring of Bethe vacua for small values of $N$, $k$ just as in the $B$ twist and find agreement with the fusion rules for the minimal models when $N = 1$.
For $N > 1$, we find that the fusion rules exactly match those found in Section \ref{sec:osp} upon exchanging $N \leftrightarrow k$.
We will provide an explanation for this in the next Section: the VOAs $L_k(\fosp_{1|2N})$ and $W^{\rm min}_{N-\scriptstyle{\frac{1}{2}}}(\fsp_{2k})$ are level-rank dual to one another.
This level-rank duality implies that the fusion rules for the minimal $W$-algebra are identical to those of the $\fosp_{1|2N}$ current algebra.

\section{Mirror symmetry and level-rank duality}
\label{sec:mirrorBC}

We argued in Section \ref{sec:osp} that the theory $\CT^B_{N,k}$ admits a boundary condition $\textnormal{Dir}_{N,k}$ that furnishes the affine VOA $L_k(\fosp_{1|2N})$ and proposed a realization of modules thereof in terms of bulk line operators ending on this boundary condition.
Similarly, in Section \ref{sec:Wmin} we argued for the existence of the minimal $W$-algebra $W^{\textnormal{min}}_{k-\scriptstyle{\frac{1}{2}}}(\fsp_{2N})$ at the boundary of $\CT_{N,k}^A$.
In this section we describe the mirror boundary conditions of $\ol{\CT}_{N,k} = (\CT_{k,N})^\vee$.
Using the observation in Section \ref{sec:3dmirror} that $\ol{\CT}_{N,k}$ differs from $\CT_{N,k}$ by a parity transformation, we are lead to a novel level-rank duality of VOAs in Section \ref{sec:sandwich} that we prove in Section \ref{sec:level-rank}.

\subsection{The boundary conditions $\textnormal{Neu}_{N,k}$ and $\textnormal{Neu}'_{N,k}$}

We now describe two boundary conditions of the UV theory introduced in Section \ref{sec:3dmirror} that we expect will flow to holomorphic boundary conditions of $\ol{\CT}_{N,k}$ that are deformable to its $A$ and $B$ twists.

\subsubsection{$N = k = 1$}
To motivate our proposal, we start by considering a boundary condition analogous to the one in \cite[Section 4.4]{Ferrari:2023fez}.
The UV boundary condition imposes Neumann boundary conditions on both the vector multiplet and the chiral multiplet.
Such a boundary condition is ill-posed by itself due to a gauge anomaly and must be supplemented by auxiliary 2d degrees of freedom; the contribution of the bulk fields to the (pure gauge) anomaly is $-2$, which can be canceled by introducing $2$ boundary Fermi multiplets $\gamma_{b}, \overline{\gamma}^{b}$, $b = 1,2$, with the action of gauge transformations generated by the current
\be
	\sigma = :\!\gamma_{1}\overline{\gamma}^{1}\!: + :\!\gamma_{2}\overline{\gamma}^{2}\!:\,.
\ee
The algebra of boundary local operators in the $HT$ twist of the UV theory is given by the (derived) gauge invariants of the algebra freely generated by a bosonic field $\phi$ of charge $1$ -- the boundary value of the chiral multiplet scalar field -- and the boundary fermions $\gamma_b, \overline{\gamma}^b$; the bosonic field $\phi$ has regular OPEs with all of the fermions, \emph{cf.} \cite[Section 6]{Costello:2020ndc}.

These (derived) gauge invariants are computed by introducing a fermionic field $c$, which also has regular OPEs with all other generators, together with a differential
\be
	Q c = 0 \qquad Q \phi = c \phi \qquad Q \gamma_b = c \gamma_b \qquad Q \overline{\gamma}_b = -c \overline{\gamma}_b
\ee
which is simply the Chevalley-Eilenberg complex for the action of the Lie algebra $\C[\![z]\!]$ of gauge transformations.
The (derived) gauge invariants are then realized as the $Q$-cohomology of the chargeless subalgebra generated by $\phi$, $\pd c$, $\gamma_b$, and $\overline{\gamma}^b$, mathematically this is simply to taking relative Lie algebra cohomology.
We find that these invariants are generated by the bosonic operators
\be
	\wt{h} = :\!\overline{\gamma}^1 \gamma_1\!: - :\!\overline{\gamma}^2 \gamma_2\!: \qquad \wt{e} = :\!\overline{\gamma}^1 \gamma_2\!: \qquad \wt{f} = :\! \overline{\gamma}^2 \gamma_1\!:\,,
\ee
realizing a copy of $L_1(\mathfrak{sl}_2)$, together with the fermionic operators
\be
	\wt{\theta}_+ = :\! \phi \overline{\gamma}^1\!: \qquad \wt{\theta}_- = :\!\phi \overline{\gamma}^2\!:\,,
\ee
transforming in the fundamental representation of these $\fsl_2$ currents.
The fermionic generators $\wt{\theta}_\pm$ have regular OPEs with themselves and with one another.
In fact, we see that this is precisely the same vertex algebra found in \cite[Section 4.1]{Ferrari:2023fez} on Dir${}_{1,1}$ prior to passing to the $B$ twist of $\CT_{1,1}$.

As on Dir${}_{1,1}$, this vertex algebra should have two natural deformations compatible with deforming the bulk to its $A$ and $B$ twists -- one deformation leads to $L_1(\fosp_{1|2})$ and the other to $M(2,5)$.
We note that $:\! \wt{\theta}_- \wt{\theta}_+ \!: = :\! \phi^2 \overline{\gamma}^1 \overline{\gamma}^2\!:$ has regular OPEs with all of the generators and hence should arise as a bulk local operator brought to the boundary \cite{Costello:2020ndc, Zeng:2021zef}.
In fact, it is the boundary value of the bulk local operator $\phi^2 V_{1}$ used to deform (the $HT$ twist of) $\ol{\CT}_{1,1}$ to its $B$ twist.
On the other hand, as the bulk fermion vanishes on this boundary condition, the operator used to deform (the $HT$ twist of) $\ol{\CT}_{1,1}$ to its $A$ twist vanishes on the boundary.
Together, these imply that Neu${}_{1,1}$ is deformable to the $A$-twist, but not the $B$-twist.

The topological flavor symmetry present on Dir${}_{1,1}$ is identified with the outer%
\footnote{Although this is an inner symmetry of the fermions, generated by the current $\sigma' = :\! \gamma_{1} \overline{\gamma}^1 \!:$, the mixed anomaly $\sigma(z) \sigma'(w) \sim (z-w)^{-2}$ implies $Q \sigma' = \partial c$, whence only $\sigma'_{-1}$ is $Q$-closed.} %
symmetry that scales $\gamma_1$, $\overline{\gamma}^1$ with weights $1,-1$.
The half-index counting local operators on this boundary condition is thus given by
\be
	\wh{I\!\!I}_{1,1}(y, \eta;q) = (q)_\infty \oint \frac{\textnormal{d} z}{2\pi i \, z} \frac{FF(-q^{\scriptstyle{\frac{1}{2}}} z y^{-1} \eta)FF(z y)}{(z)_\infty}
\ee
where $FF(x) = (x)_\infty(q x^{-1})_\infty$.
In this expression, $y$ is the fugacity/Jacobi variable for the Cartan subalgebra generated by $\wt{h}$ and $\eta$ is the fugacity for the symmetry rotating $\gamma_1$, $\overline{\gamma}^1$; $z$ is again the fugacity for gauge transformations.
It is a simple application of the bosonization formula
\be
	FF(x) = \sum_{m}(-1)^m \frac{q^{\scriptstyle{\frac{1}{2}}m(m-1)} x^m}{(q)_\infty}
\ee
and the $q$-binomial identity
\be
	\frac{1}{(x)_\infty} = \sum_{l \geq 0}\frac{x^l}{(q)_l}
\ee
to show that $\wh{I\!\!I}_{1,1}(y,\eta;q) = \wt{I\!\!I}_{1,1}(y, \eta^{-1};q)$.
In particular, the $A$-twist specialization $\eta \to - q^{-\scriptstyle{\frac{1}{2}}}$ agrees with the (normalized) vacuum character of $L_1(\fosp_{1|2})$
\be
	\ol{I\!\!I}^A_{1,1}(y;q) = \wh{I\!\!I}_{1,1}(y,-q^{-\scriptstyle{\frac{1}{2}}};q) = \ol{\chi}[L_1(\fosp_{1|2})](y;q)\,.
\ee
Putting this together, we are lead to conclude the boundary condition Neu${}_{1,1}$ of $\ol{\CT}_{1,1}$ is the 3d mirror of the boundary condition Dir${}_{1,1}$ of $\CT_{1,1}$.

One immediate conclusion from this analysis is that the boundary condition Neu${}_{1,1}$ must be modified to be compatible with the $B$ twist --- the operator $\phi^2 V_{1}$ used to deform $\ol{\CT}_{1,1}$ to its $B$ twist does not vanish on Neu${}_{1,1}$ and so this boundary condition is not deformable to the $B$ twist \cite{Brunner:2021tfl, Ferrari:2023fez}.
This directly mirrors the deformation from Dir${}_{1,1}$ to Dir${}'_{1,1}$, the former being compatible with the $B$ twist of $\CT_{1,1}$ and the latter with the $A$ twist.
Nonetheless, the half-index counting local operators in the $B$ twist of $\ol{\CT}_{1,1}$ can be realized as in Section 4.4 of \cite{Ferrari:2023fez} by also specializing $y$ and agrees with the (normalized) vacuum character of $M(2,5)$
\be
	\ol{I\!\!I}^B_{1,1}(q) = \wh{I\!\!I}_{1,1}(1,-q^{\scriptstyle{\frac{1}{2}}};q) = \ol{\chi}[M(2,5)](q)\,.
\ee

\subsubsection{General $N$, $k$}

We now extend this to a boundary condition of $\ol{\CT}_{N,k}$ for general $N$, $k$.
We start by imposing Neumann boundary conditions on both the vector multiplets and chiral multiplets.
We must again introduce auxiliary degrees of freedom to avoid gauge anomalies; the contribution from the bulk to the gauge anomaly is given by the negative of $\kappa$, which we remedy with a certain choice of boundary Fermi multiplets.
Explicitly, we introduce $2Nk$ boundary Fermi multiplets $\gamma_{jb}, \overline{\gamma}^{jb}$, where $j = 1, \dots, k$ and $b = 1, \dots, 2N$.
We take the action of gauge transformations on these fermions to be generated by the currents
\be
	\sigma_{ia} = (-1)^{a+1}\sum_{j=1}^i :\!\gamma_{ja}\overline{\gamma}^{ja}\!: + :\!\gamma_{j(a+1)}\overline{\gamma}^{j(a+1)}\!:
\ee
which indeed have the requisite level to cancel the anomaly
\be
	\sigma_{I}(z) \sigma_{J}(w) \sim \frac{\kappa_{IJ}}{(z-w)^2}\,.
\ee
The $k$ currents
\be
	\wt{h}_i = \sum_{b=1}^{2N} (-1)^{b+1} :\!\gamma_{ib}\overline{\gamma}^{ib}\!:
\ee
each have level $2N$ and commute with one another and with the $\sigma_{ia}$.

The half-index counting local operators on the left boundary condition $\textnormal{Neu}_{N,k}$ takes the form
\be
\begin{aligned}
	\wh{I\!\!I}_{N,k}(y_i, \eta;q) & = \oint \prod\limits_{i=1}^k \prod\limits_{a=1}^{2N-1}\bigg(\frac{\textnormal{d} z_{ia}}{2\pi i \, z_{ia}} \frac{(q)_\infty}{(z_{ia})_\infty} \bigg)\\
	& \qquad \times \prod\limits_{i=1}^{k}\prod\limits_{\alpha=1}^{N}\bigg( FF\big(y_i \prod\limits_{j=i}^k \frac{z_{j (2\alpha-1)}}{z_{j(2\alpha-2)}}\big) FF\big(-q^{\scriptstyle{{\frac{1}{2}}}} \eta y_i^{-1} \prod\limits_{j=i}^k \frac{z_{j (2\alpha-1)}}{z_{j(2\alpha)}}\big)\bigg)\\
\end{aligned}
\ee
where the integration contour is taken to be the product of unit circles $|z_{ia}|=1$ and we set $z_{i 0} = z_{i 2N} = 1$.
The contributions in the first line come from the bulk fields restricted to the boundary, whereas the second line realizes contributions from the boundary fermions.

In order to count local operators in the $A$ and $B$ twists, we must specialize this half-index as we did for $N = k = 1$ above. In particular, the half-index counting local operators in the $A$ twist is given by
\be
	\ol{I\!\!I}^A_{N,k}(y_i;q) = \wh{I\!\!I}_{N,k}(y_i, \eta;q)\bigg|_{(\diamondsuit)}\,.
\ee
where $(\diamondsuit)$ denotes specializing $\eta \to -q^{-\scriptstyle{{\frac{1}{2}}}}$.
We expect that this agrees with the (normalized) vacuum character of $L_N(\fosp_{1|2k})$:
\be
	\ol{I\!\!I}^A_{N,k}(y_i;q) \overset{?}{=} \ol{\chi}[L_N(\fosp_{1|2k})](y_i; q)
\ee
We have checked this expectation is true for all $N$, $k$ with $N+k \leq 4$ to order at least $q^4$.

Similarly, the half-index counting local operators in the $B$ twist is realized by specializing $\eta \to -q^{\scriptstyle{{\frac{1}{2}}}}$.
We also expect that the $B$-twist deformation will force us to specialize $y_1 \to 1$, and we set $u_l = y_{l + 1}$ for $l = 1, \dots, k-1$:
\be
 	\ol{I\!\!I}^B_{N,k}(u_l;q) = \wh{I\!\!I}_{N,k}(y_i, \eta;q)\bigg|_{(\diamondsuit\diamondsuit)}
\ee
where $(\diamondsuit \diamondsuit)$ denotes specializing $y_1 \to 1$, $y_{l+1} \to u_l$ for $l = 1, \dots k-1$, and $\eta \to -q^{\scriptstyle{{\frac{1}{2}}}}$.
We have again checked that this half-index reproduces the (normalized) vacuum character of $W^{\textnormal{min}}_{N- \scriptstyle{\frac{1}{2}}}(\fsp_{2k})$ to order at least $q^4$ for all $N$, $k$ satisfying $N+k \leq 4$, leading to the expectation 
\be
	\ol{I\!\!I}^B_{N,k}(q^{\scriptstyle{\frac{1}{2}}} u_l;q) \overset{?}{=} \ol{\chi}[W^{\textnormal{min}}_{N - \scriptstyle{\frac{1}{2}}}(\fsp_{2k})](u_l; q)\,.
\ee

Just as in Sections \ref{sec:osp} and \ref{sec:Wmin}, we expect that all modules for these boundary VOAs are realized by the local operators living at the end of bulk line operators.
The half-index counting local operators at the end of such a line operator $\CL$ takes the form
\be
\begin{aligned}
	\wh{I\!\!I}_{N,k}[\CL](y_i, \eta;q) & = \oint \prod\limits_{i=1}^k \prod\limits_{a=1}^{2N-1}\bigg(\frac{\textnormal{d} z_{ia}}{2\pi i \, z_{ia}} \frac{(q)_\infty}{(z_{ia})_\infty} \bigg) \CP(z^{-1})\\
	& \qquad \times \prod\limits_{i=1}^{k}\prod\limits_{\alpha=1}^{N}\bigg( FF\big(y_i \prod\limits_{j=i}^k \frac{z_{j (2\alpha-1)}}{z_{j(2\alpha-2)}}\big) FF\big(-q^{\scriptstyle{\frac{1}{2}}}\eta y_i^{-1} \prod\limits_{j=i}^k \frac{z_{j (2\alpha-1)}}{z_{j(2\alpha)}}\big)\bigg)\\
\end{aligned}
\ee
which we further specialize via $(\diamondsuit)$ and $(\diamondsuit \diamondsuit)$ to count local operators in the $A$ and $B$ twists, respectively.

We expect that the line operators $\ol{\CL}^A_\mu$ and $\ol{\CL}^B_\mu$ identified in Section \ref{sec:3dmirror} reproduce all simple modules of the corresponding VOAs.
The half-indices counting local operators are given by
\be
\ol{I\!\!I}^A_{N,k}[\ol{\CL}^A_\mu](y_i;q) = \wh{I\!\!I}_{N,k}[\ol{\CL}^A_\mu](y_i, \eta;q)\bigg|_{(\diamondsuit)}
\ee
in the $A$ twist and
\be
\ol{I\!\!I}^B_{N,k}[\ol{\CL}^B_\mu](u_l;q) = \wh{I\!\!I}_{N,k}[\ol{\CL}^B_\mu](y_i, \eta;q)\bigg|_{(\diamondsuit \diamondsuit)}
\ee
in the $B$ twist.
We have checked that these half-indices reproduce the characters of all simple modules for $L_N(\fosp_{1|2k})$ and $W^{\textnormal{min}}_{N - \scriptstyle{\frac{1}{2}}}(\fsp_{2k})$, respectively, for $N + k \leq 4$ to order at least $q^4$.

\subsection{Mirror symmetry and sandwiches}
\label{sec:sandwich}
We observed in Section \ref{sec:3dmirror} that $\ol{\CT}_{N,k}$ (the 3d mirror of $\CT_{k,N}$) is simply the parity conjugate of $\CT_{N,k}$.
As a result, we can equivalently think of the above \emph{right} boundary conditions of $\ol{\CT}_{N,k}$ as \emph{left} boundary conditions of $\CT_{N,k}$.
Correspondingly, we find that the TQFT $\CT^B_{N,k}$ admits two boundary VOAs modeling its categories of line operators, one on the left and one on the right: it furnishes
\be
	\CV^B_{N,k} = L_{k}(\fosp_{1|2N})
\ee
on the right and 	
\be
	{}^!\CV^B_{N,k} = \CV^A_{k,N} = W^{\textnormal{min}}_{N-\scriptstyle{\frac{1}{2}}}(\fsp_{2k}) 
\ee
on the left.
The same is true for $\CT^A_{N,k}$: it furnishes
\be
	\CV^A_{N,k} = W^{\textnormal{min}}_{k-\scriptstyle{\frac{1}{2}}}(\fsp_{2N})	\quad \text{and} \quad {}^!\CV^A_{N,k} = \CV^B_{k,N} = L_N(\fosp_{1|2k})\,.
\ee
Both the left VOA and the right VOA can be used to model the category of bulk line operators, hence their categories of modules should be equivalent to one another, both being equivalent to the category of bulk line operators.
This equivalence of module categories is induced by identifying the modules living at the ends of the same bulk line operator, \emph{cf.} Fig. \ref{fig:sandwich}.
We note that the braiding on the category of modules for the left VOA must be reversed in order to capture the relative orientations of the two boundaries.
Based on the analogy with ordinary Chern-Simons gauge theories, the resulting braid-reversed equivalence of boundary VOAs can be thought of as a version of level-rank duality.
\be
	\textnormal{level-rank duality: } L_k(\fosp_{1|2N}) \overset{?}{\longleftrightarrow} W^{\textnormal{min}}_{N-\scriptstyle{\frac{1}{2}}}(\fsp_{2k})
\ee
See \cite[Section 4.4]{Ferrari:2023fez} for the case $N = k = 1$, where one finds the minimal model $M(2,5)$ on the left of $\CT^B_{1,1}$.

It is important to note that we can place these TQFTs on a spacetime of the form $\Sigma \times [0,1]$ with these left and right holomorphic boundary conditions placed at $\Sigma \times 0$ and $\Sigma \times 1$, resulting in an effective 2d holomorphic theory on $\Sigma$.
The local degrees of freedom in this effective 2d theory are realized by the local operators on the two boundaries together with all line operators stretched between the two boundaries.
The character that counts these local operators comes from gluing our half-indices together.
\be
\begin{aligned}
	I^{A}_{2\textnormal{d}}(y_i, w_\nu; q) = \sum\limits_{\mu} \ol{I\!\!I}^{A}_{N,k}[\CL^{A}_\mu](y_i;q)  I\!\!I^A_{N,k}[\CL^{A}_\mu] (w_\nu;q)\\
	I^{B}_{2\textnormal{d}}(u_l, x_\alpha;q) = \sum\limits_{\mu} \ol{I\!\!I}^{B}_{N,k}[\CL^{B}_\mu](u_l;q) I\!\!I^B_{N,k}[\CL^{B}_\mu](x_\alpha;q)
\end{aligned}
\ee
The term with $\mu = 0$ counts the local operators in the effective 2d theory coming from the two boundaries.
The remaining terms count all possible local operators sitting at the two ends of the remaining simple line operators.
Based on series expansions for $N + k \leq 4$, we expect that this sum (after performing the shift $w_\nu \to q^{\scriptstyle{\frac{1}{2}}} w_\nu$) is equal to
\be
	I^A_{2\textnormal{d}}(y_i, q^{\scriptstyle{\frac{1}{2}}} w_\nu;q) \overset{?}{=} \prod\limits_{i=1}^{k}\bigg(FF\big(y_i\big) FF\big(y_i^{-1}\big) \prod\limits_{\nu=1}^{N-1} F\big(q^{\scriptstyle{\frac{1}{2}}}y_i w_\nu\big) F\big(q^{\scriptstyle{\frac{1}{2}}}y_i^{-1} w_\nu\big) \bigg)
\ee
in the $A$ twist, which can be identified with the vacuum character of a $bc$ ghost system transforming in the fundamental representation of $\fsp_{2k}$ times free fermions transforming in the fundamental representation of $\fsp_{2k} \oplus \fsp_{2(N-1)}$.
Similarly, in the $B$ twist we find
\be
	I^B_{2\textnormal{d}}(q^{\scriptstyle{\frac{1}{2}}} u_l, x_\alpha;q) \overset{?}{=} \prod\limits_{\alpha=1}^{N}\bigg(FF\big(x_\alpha\big) FF\big(x_\alpha^{-1}\big) \prod\limits_{l=1}^{k-1} F\big(q^{\scriptstyle{\frac{1}{2}}}x_\alpha u_l\big) F\big(q^{\scriptstyle{\frac{1}{2}}}x_\alpha^{-1} u_l\big) \bigg)
\ee
which can be identified with the vacuum character of a $bc$ ghost system transforming in the fundamental representation of $\fsp_{2N}$ times free fermions transforming in the fundamental representation of $\fsp_{2N} \oplus \fsp_{2(k-1)}$.

\subsection{Level-rank duality of $L_k(\mathfrak{osp}_{1|2N})$ and $W^{\textnormal{min}}_{N-\scriptstyle{\frac{1}{2}}}(\mathfrak{sp}_{2k})$}\label{sec:level-rank}

In the previous subsections, we argued that each of the TQFTs $\CT^A_{N,k}$ and $\CT^B_{N,k}$ admits both left and right boundary conditions furnishing VOAs.
From a vertex-algebraic perspective, these observations for $\CT^B_{N,k}$ suggest that the categories of modules for
\be
	L_k(\fosp_{1|2N}) \qquad \textnormal{and} \qquad W^{\textnormal{min}}_{N-\scriptstyle{\frac{1}{2}}}(\fsp_{2k})
\ee
are equivalent as braided tensor categories, up to a reversal of the braiding.
We now give a proof of this braid-reversed equivalence: these two VOAs form a commuting pair inside a free-fermion VOA.

Let $F^N$ denote the VOA of $N$ free, real fermions and let $\beta\gamma^m$ denote $m$ $\beta\gamma$-ghost VOAs, also often called symplectic bosons.
We start by considering $F^{4Nk} \otimes \beta\gamma^k$.
This VOA should be thought of as the free field algebra in the standard representation of the Lie algebra $\mathfrak{osp}_{1|2N} \oplus \mathfrak{sp}_{2k}$.
In particular, this VOA admits a conformal embedding of $\widetilde V_k(\mathfrak{osp}_{1|2N}) \otimes \widetilde V_{N-\scriptstyle{\frac{1}{2}}}(\mathfrak{sp}_{2k})$.
Here $\widetilde V$ is a homomorphic image of the universal affine VOA.
\emph{A priori} they need not be simple quotients, but in our cases it is due to the following facts.
Corollary 2.1 of \cite{Creutzig:2022sfx} tells us that $\widetilde V_k(\mathfrak{osp}_{1|2N}) = L_k(\mathfrak{osp}_{1|2N})$.
Its coset in $F^{4Nk} \otimes \beta\gamma^k$ is a conformal extension of $\widetilde V_{N-\scriptstyle{\frac{1}{2}}}(\mathfrak{sp}_{2k})$ and this extension must be simple by \cite{ACK}.
Thus by Theorem 3.5 of \cite{Arakawa:2021ogm} also 
$\widetilde V_{N-\scriptstyle{\frac{1}{2}}}(\mathfrak{sp}_{2k}) \cong L_{N-\scriptstyle{\frac{1}{2}}}(\mathfrak{sp}_{2k})$.
Next, one would like to know if these two vertex algebras form a mutually commuting pair.
Theorem 6.3 of \cite{LS1} gives us that indeed $\textnormal{Com}\left(L_{N-\scriptstyle{\frac{1}{2}}}(\mathfrak{sp}_{2k}), F^{4Nk} \otimes \beta\gamma^k \right)\cong L_k(\mathfrak{osp}_{1|2N})$. 

We can understand our desired level-rank duality by Hamiltonian reduction.
Denote by $H^{\textnormal{min}}_\ell$ the minimal Hamiltonian reduction functor on the category of modules of the affine vertex algebra $L_\ell(\mathfrak{sp}_{2k})$.
In order to understand the reduction of $F^{4Nk} \otimes \beta\gamma^k$, we note that there is an action of $L_{N}(\mathfrak{sp}_{2k})$ on the first factor, an action of $L_{-\scriptstyle{\frac{1}{2}}}(\mathfrak{sp}_{2k})$ on the second one, and a diagonal action of $L_{N-\scriptstyle{\frac{1}{2}}}(\mathfrak{sp}_{2k})$.
The Main Theorem of \cite{Arakawa:2020oqo} applied to this case implies
\be
	H^{\textnormal{min}}_{N-\scriptstyle{\frac{1}{2}}}\left( F^{4Nk} \otimes \beta\gamma^k\right) \cong 
	F^{4Nk} \otimes H^{\textnormal{min}}_{-\scriptstyle{\frac{1}{2}}}\left(  \beta\gamma^k\right).
\ee
This isomorphism preserves conformal vectors if we view $4N$ of these free fermions as $2N$ $bc$-ghost systems of conformal weight $(1, 0)$, while all others have the usual conformal vector that gives them conformal weight $\frac{1}{2}$. 
Now, $H^{\textnormal{min}}_{-\scriptstyle{\frac{1}{2}}}\left(  \beta\gamma^k\right) \cong \mathbb C^2$ by Lemma 7.1 and Remark 7.3 of \cite{Creutzig:2021dda}.
The factor of $\mathbb C^2$ appears because the minimal reduction functor for $\mathfrak{sp}_{2k}$ is two-to-one at levels with denominator two.
As the reduction functor $H^{\textnormal{min}}_\ell$ maps simple modules to simple modules when $\ell$ is not a positive integer \cite{Ara05}, we conclude 
$L_k(\mathfrak{osp}_{1|2N})$ and $W^{\textnormal{min}}_{N-\scriptstyle{\frac{1}{2}}}(\mathfrak{sp}_{2k})$ form a commuting pair inside $F^{4Nk}$, but not necessarily that they are mutual commutants.

To establish that they are mutual commutants, we study the branching rules of $F^{4Nk}$.
We denote the simple highest-weight module of highest-weight $\lambda$ of the affine vertex algebra at level $k$ by the symbol $L_k(\lambda)$.
For this we first consider $L_{k-1}(\mathfrak{sp}_{2N}) \otimes F^{4N}$.
In \cite{Creutzig:2024uid} it was shown that there is in fact a conformal embedding of $L_k(\mathfrak{osp}_{1|2N})$ and its coset is the principal $W$-algebra $W_\ell(\mathfrak{sp}_{2N})$ of $\mathfrak{sp}_{2N}$ at level $\ell = -(N+1) + \frac{N+k}{2(N+k)+1}$.
It has simple modules that are denoted by $L_\ell(\mu, \lambda)$ with $\mu, \lambda$ in certain subsets of the set of dominant weights of $\mathfrak{sp}_{2N}, \mathfrak{so}_{2N+1}$, see \cite{Arakawa:2015sei}.
In Section 9 of \cite{Creutzig:2024uid} the following decomposition formula was derived 
\be\label{eq:decomp1}
    L_{k-1}(\mu) \otimes F^{4N} = \bigoplus_{\lambda \in P^B_Q} L_k(\lambda) \otimes L_\ell(\mu, \lambda)\,.
\ee
Here the $ L_k(\lambda)$ are modules of $L_k(\mathfrak{osp}_{1|2N})$ and the set $P^B_Q$ is defined in \cite{Creutzig:2022riy}, it parametrizes the simple modules of $L_k(\mathfrak{osp}_{1|2N})$.
The symplectic level-rank duality of \cite{ostrik2020symplectic} says
\be\label{eq:decomp2}
    F^{4N(k-1)} \cong \bigoplus_{\lambda \in P^N_{k-1}} L_N(\lambda^\sigma) \otimes L_{k-1}(\lambda)
\ee
as module for $L_{k-1}(\mathfrak{sp}_{2N}) \otimes L_{N}(\mathfrak{sp}_{2(k-1)})$.
Here $P^N_{k-1}$ denotes the set of simple ordinary modules of $L_{k-1}(\mathfrak{sp}_{2N})$ and $\sigma: P^N_{k-1} \rightarrow P_N^{k-1}, \lambda \mapsto \lambda^\sigma$ is a certain bijection that in particular maps the first fundamental weight $\omega_1$ of $\mathfrak{sp}_{2N}$ to the one of $\mathfrak{sp}_{2(k-1)}$.
$F^{4N(k-1)}$ is just the vertex superalgebra of $4N(k-1)$ free fermions.
The fermions carry the standard representation of both the affine $\mathfrak{sp}_{2N}$ and of $\mathfrak{sp}_{2(k-1)}$.
Lemma 2.4 of \cite{Creutzig:2022riy} applied to this example implies that the fusion ring of $L_N(\mathfrak{sp}_{2(k-1)})$ is generated by $L_N(\omega_1)$.
Combining \eqref{eq:decomp1} and \eqref{eq:decomp2} we get
\be\label{eq:decomp3}
\begin{split}
  F^{4Nk} &\cong   F^{4N(k-1)} \otimes F^{4N} \\
  &\cong \bigoplus_{\mu \in P^N_{k-1}} L_N(\mu^\sigma) \otimes L_{k-1}(\mu) \otimes F^{4N} \\
  &\cong \bigoplus_{\mu \in P^N_{k-1}}\bigoplus_{\lambda \in P^B_Q}  L_k(\lambda) \otimes L_N(\mu^\sigma) \otimes  L_\ell(\mu, \lambda)\,.
\end{split}
\ee
Thus the coset is
\be
    \text{Com}\left(L_k(\mathfrak{osp}_{1|2N}), F^{4Nk}\right) = \bigoplus_{\mu \in P^N_{k-1}}  L_N(\mu^\sigma)  \otimes L_\ell(\mu, 0)\,.
\ee
By Theorem 7.7 of \cite{Creutzig:2021dda} $W^{\textnormal{min}}_{N-\scriptstyle{\frac{1}{2}}}(\mathfrak{sp}_{2k})$ is a conformal extension of $L_N(\mathfrak{sp}_{2(k-1)}) \otimes W_\ell(\mathfrak{sp}_{2N})$. 
As the minimal $W$-algebra is generated by its affine vertex subalgebra, the Virasoro element and the weight $3/2$ fields in the standard representation this means that this extension is given by these fields of weight $3/2$.
We thus need to verify that the coset $\text{Com}\left(L_k(\mathfrak{osp}_{1|2N}), F^{4Nk}\right)$ is also just an extension of $L_N(\mathfrak{sp}_{2(k-1)}) \otimes W_\ell(\mathfrak{sp}_{2N})$ by weight $3/2$ fields in the standard representation of $\mathfrak{sp}_{2(k-1)}$.

As the fusion ring of $L_N(\mathfrak{sp}_{2(k-1)})$ is generated by $L_N(\omega_1)$ it follows from Lemma 2.4 \cite{Creutzig:2022riy} that $\text{Com}\left(L_k(\mathfrak{osp}_{1|2N}), F^{4Nk}\right)$ is generated under OPE by $L_N(\mathfrak{sp}_{2(k-1)}) \otimes W_\ell(\mathfrak{sp}_{2N})$ together with any field of $L_N(\omega_1)  \otimes L_\ell(\omega_1, 0)$, \emph{i.e.} we can take a field corresponding to the top level.
The conformal weight of the top level of $L_N(\lambda)$ is
\be
    h(\lambda, N) = \frac{(\lambda, \lambda+2\rho)}{2(N+k)}
\ee
and the one of $L_\ell(\lambda, 0)$ is (see, \emph{e.g.} \cite[Eq. (30)]{Arakawa:2018iyk})
\be
    h^W(\lambda, \ell) = \frac{(2(N+k)+1)(\lambda, \lambda+2\rho)}{2(N+k)} - \lambda \rho^\vee
\ee
with $\rho, \rho^\vee$ the Weyl vector and the dual Weyl vector.
We compute that 
\be
    h(\omega_1, N) + h^W(\omega_1, \ell) = \frac{2k-1}{4(N+k)} + \frac{(2(N+k)+1)(2N+1)}{4(N+k)} - (N-\frac{1}{2}) = \frac{3}{2}.
\ee
Thus we can conclude that $\text{Com}\left(L_k(\mathfrak{osp}_{1|2N}), F^{4Nk}\right) = W^{\textnormal{min}}_{N-\scriptstyle{\frac{1}{2}}}(\mathfrak{sp}_{2k})$.
Set 
\be
    W^{\textnormal{min}}_{N-\scriptstyle{\frac{1}{2}}}(\lambda) := \bigoplus_{\mu \in P^N_{k-1}}  L_N(\mu^\sigma) \otimes  L_\ell(\mu, \lambda).
\ee
we note that these modules are pairwise non-isomorphic, since $L_\ell(0, \lambda) \not\cong L_\ell(0, \lambda')$ for $\lambda\neq \lambda'$, this is because the only identification is $L_{\ell}(\lambda, \mu) \cong L_{\ell}(\omega(\lambda), \omega(\lambda'))$ and $\omega$ is the affine Weyl translation corresponding to the $n$-th fundamental weight $\omega_n$ and this translations only maps $0$ to itself at boundary admissible levels. 
Thus
\be\label{eq:decomp4}
\begin{split}
  F^{4Nk} &\cong \bigoplus_{\lambda \in P^B_Q}  L_k(\lambda) \otimes W^{\textnormal{min}}_{N-\scriptstyle{\frac{1}{2}}}(\lambda)
\end{split}
\ee

This means that  there is a braid-reversed equivalence $\tau$ of braided tensor supercategories \cite{Creutzig:2019psu, mcrae2023general}, such that
\be\label{eq:level-rank}
	\tau( L_k(\lambda)) = W^{\textnormal{min}}_{N-\scriptstyle{\frac{1}{2}}}(\lambda)^*
\ee
The upper index ${}^*$ indicates the dual. In particular the fusion rings of both categories are the same.

\bibliography{osp}
\bibliographystyle{amsalpha}

\end{document}